\documentclass[a4paper,12pt]{article}

\usepackage[pdftex]{graphicx}  
\usepackage[pdftex]{graphics}  

\usepackage{amsmath}
\usepackage{amssymb}
\usepackage{amsfonts}
\usepackage{theorem}
\usepackage{bbm}
\usepackage{varioref}

\numberwithin{equation}{section}

\theorembodyfont{\slshape}  
\newtheorem{theo}{Theorem}[section]
\newtheorem{coro}[theo]{Corollary}
\newtheorem{prop}[theo]{Proposition}
\newtheorem{defi}[theo]{Definition}
\newtheorem{lemm}[theo]{Lemma}
\newtheorem{rema}[theo]{Remark}

\newtheorem{obse}[theo]{Observation}
\newtheorem{summ}[theo]{Summary}
\newtheorem{note}[theo]{Notational Remark}

\newtheorem{proc}[theo]{The Procedure}

\newenvironment{proof}
{\begin{trivlist}\item[]\textsc{Proof}.\begin{small}}
{\end{small}%
\hspace*{\fill}\nolinebreak[1]\hspace*{\fill}\rule{1.5ex}{1.5ex}%
\end{trivlist}}

\newenvironment{prooff}
{\begin{trivlist}\item[]\textsc{Proof}.}
{%
\hspace*{\fill}\nolinebreak[1]\hspace*{\fill}\rule{1.5ex}{1.5ex}%
\end{trivlist}}

\newenvironment{skproof}
{\begin{trivlist}\item[]\textsc{Sketch of Proof}.\begin{small}}
{\end{small}%
\hspace*{\fill}\nolinebreak[1]\hspace*{\fill}\rule{1.5ex}{1.5ex}%
\end{trivlist}}

\newcommand{\betr}[1]{| #1 |}
\newcommand{\norm}[1]{\| #1 \|}
\newcommand{\skal}[2]{\langle #1 | #2 \rangle}

\newcommand{\ovl}[1]{\overline{#1}}     

\newcommand{\ra}{\rightarrow}           

\newcommand{\cH}{\mathcal{H}}

\newcommand{\bB}{\mathbf{B}}

\newcommand{\bE}{\mathbf{E}}

\newcommand{\dA}{{\mathbb{A}}}

\newcommand{\dC}{{\mathbbm{C}}}

\newcommand{\dN}{{\mathbbm{N}}}

\newcommand{\dR}{{\mathbbm{R}}}

\newcommand{\eps}{\epsilon}

\newcommand{\vp}{\varphi}

\newcommand{\om}{\omega}

\newcommand{\Lam}{\Lambda}

\newcommand{\e}{{\operatorname{e}}}
\newcommand{\ex}[1]{\operatorname{exp}\{ #1 \}}

\newcommand{\dom}{\operatorname{dom}}        
\newcommand{\supp}{\operatorname{supp}}      
\renewcommand{\div}{\operatorname{div}}      
\newcommand{\curl}{\operatorname{curl}}      

\newcommand{\set}[2]{\{ #1 \,|\, #2 \}}

\usepackage{vmargin}
\setpapersize{A4}
\setmargins{28mm}{25mm}{155mm}{228mm}{\baselineskip}{5.5ex}{0mm}{0mm}

\IfFileExists{fancyhdr.sty}{\usepackage{fancyhdr}}%
{\usepackage{fancyheadings}}
\begin{document}
\begin{sloppypar}

%
%

%
%
\title{On Wave-Like Differential Equations\\
          in General Hilbert Space.\\
      The Functional Analytic Investigation of\\ 
      Euler--Bernoulli Bending Vibrations of a Beam\\
      as an Application in Engineering Science.}

\author{Reinhard Honegger\footnote{Reinhard.Honegger@reutlingen-university.de}, Michael Lauxmann,\footnote{Michael.Lauxmann@reutlingen-university.de}\\
and Barbara Priwitzer\footnote{Barbara.Priwitzer@reutlingen-university.de}
}

\date{February, 2024} 

\maketitle
\thispagestyle{empty}

\begin{abstract}
Wave-like partial differential equations occur in many engineering applications. Here the engineering setup is embedded into the Hilbert space framework of functional analysis of modern mathematical physics.
The notion \emph{wave-like} is a generalization of the primary wave (partial) differential equation.

A short overview over three wave-like problems in physics and engineering is presented.
The mathematical procedure for achieving positive, selfadjoint differential operators in an $\mathrm{L}^2$--Hilbert space is described, operators which then may be taken for wave-like differential equations. 
Also some general results from the functional analytic literature are summarized.

The main part concerns the investigation of the free Euler--Bernoulli bending vibrations of a slender, straight, elastic beam in one spatial dimension 
in the $\mathrm{L}^2$--Hilbert space setup. 
Taking suitable Sobolev spaces we perform the mathematically exact introduction and analysis 
of the corresponding (spatial) positive, selfadjoint differential operators of $4$-th order, which belong
to the different boundary conditions arising as supports in statics. 
A comparison with free wave swinging of a string is added, using a Laplacian as differential operator.


\end{abstract}

%
%
\tableofcontents    

\vspace{6ex}
Before starting let us remark that 
the present article is an extended, longer, more detailed, better and easier understandable version of the published work \cite{HLP24} from the same authors.


%
%
\section{Introduction, Overview, \\Definition of Wave-Like Equation}
\label{s0-introduction}
In many engineering investigations and applications wave-like partial differential equations play an important role.
The engineering calculations to solve such differential equations seem to be very specific to the specially chosen situations.
But in this way, deeper mathematical questions remain unanswered.
Nevertheless the directly calculating methods used are successful and appropriate for the selected application, which lead to concrete solutions,
analytically and numerically, e.g.~\cite{ReismannPawlik74,MaPoSe,Strauss92/08,DraHol07}.

For beam dynamics there exist four engineering theories: Euler--Bernoulli model, Rayleigh model, shear model and Timoshenk model. Classically the dynamics of the transversally bending beam is investigated and directly computed by eigenfunction expansion \cite{HanBenaroyWei99}.

Here we investigate free Euler--Bernoulli bending vibrations of a slender, straight, elastic beam
in a completely different and much larger context, namely in terms of Hilbert space methods of modern mathematical physics.

Functional analysis is capable to provide general statements for very general
situations, namely predictions on existence and smoothness degrees of eigenfunctions
and solutions (regularity). The mentioned specific engineering techniques are
far from being able to deal with this generality. 
Nevertheless there are limitations, since 
also in a Hilbert space setting it is not possible to calculate analytically eigenvalues and eigenfunctions of differential operators
on an arbitrary spatial region $\Lam\subseteq\dR^r$, $r=1,2,3$, which does not possess specific geometric properties like symmetries.

The present article aims to
bring together computational and theoretical engineering science with functional
analytic methods. The novelty of the present approach is the complete incorporation and investigation of the Euler--Bernoulli differential equation in the general context of
Hilbert space operator theory in functional analysis.

In order to be precise let us introduce, what will be understood under a wave-like differential equation in Hilbert space language.
\begin{defi}[Wave-Like Differential Equation]
\label{defi_wave-like}
A differential equation of type
\begin{equation*}
\frac{d^2u(t)}{dt^2}=-Au(t)
\end{equation*}
in a Hilbert space $\cH$ with some positive, selfadjoint operator $A$ is called to be \emph{wave-like}.
A solution of which is a trajectory $\dR\ni t\mapsto u(t)\in\cH$, where in applications the variable $t\in\dR$ is interpreted 
as the time parameter for evolution in time.
\end{defi}
In engineering or physical applications, the operators $A$ usually represent differential operators of second or higher order 
acting in some $\mathrm{L}^2$--Hilbert space of square integrable functions on a spatial region $\Lam$.
The primary wave equation concerns a Laplace operator $A=-\Delta$, so the notion \emph{wave-like} is its generalization. 
To free Euler--Bernoulli bending vibrations of a beam there belong differential operators $A$ of $4$-th order. 

In the current article we first outline the general Hilbert space solution of wave-like differential equations, see
Section~\ref{s1-general-wave}.

In Section~\ref{s2-Weq-in-L2}  the necessary mathematical procedure for obtaining positive, selfadjoint differential operators $A$ in the $\mathrm{L}^2$--Hilbert space approach is described, but also some general results from the literature are reported.
Moreover,  a short overview over three wave-like equations for arbitrary spatial regions $\Lam$ is given:
the primary wave equation as a partial differential equation,
the wave decoupling for Maxwell radiation in electromagnetism, and free bending vibrations of a plate.

Section~\ref{s3-sobolev+diffOps} is auxiliary but necessary for bending vibrations of a beam in the next Section~\ref{s3-bending-vibs}. 
It is dedicated to Sobolev spaces corresponding to diverse boundary conditions with associated differential operators of first and second order 
acting in the open bounded interval $(0,\ell)$ with $\ell>0$. 

In Section~\ref{s3-bending-vibs} we outline in detail the $\mathrm{L}^2$--Hilbert space frame in one spatial dimension
for positive, selfadjoint differential operators $A$ of $4$-th order, which describe via wave-like equations 
the free Euler--Bernoulli bending vibrations of a slender, straight, elastic beam placed in $(0,\ell)$.
At the ends of the beam, $x=0$ and $x=\ell$, the diverse support possibilities from engineering statics are taken, 
namely the combinations of flexible or fixed support, and free end.
Intrinsically involved into the domain of definition of such a differential operator $A$ of $4$-th order is
the chosen boundary condition (= support) of the beam. 
Each $A$ is identified as the well known Friedrichs extension \cite{Kato84,Weidmann} of the suitable product operator 
of four differential operators of first order respecting exactly the selected support of the beam. 
For such positive, selfadjoint $A$ we prove the existence of a purely discrete spectrum with help of a Sobolev compact embedding theorem.
Two groups of these operators $A$ are distinguished. One group with analytically solvable eigenequations, and the other group for which only numerical solutions of the eigenequations are possible, both being considered in detail in Section \ref{s4b-eigenequation}. 
That reflects directly some properties of the mentioned Friedrichs extensions. 
We derive operator properties and interrelations, which seem to be unknown in the engineering and mathematical literature. 

In Section~\ref{s4-Weq-EB-wave} we discuss similarities and differences
between free bending vibrations of a beam and free wave swinging of a string, for comparable boundary conditions.

The detailed mathematical proofs for Section~\ref{s3-bending-vibs} are given in the last Section \ref{s6-proofs}.

In order to understand the results and their proofs, the reader should be familiar with
some basics on unbounded operators and sesquilinear forms acting in Hilbert spaces,
such as closure, closedness, graph norm, core, selfadjointness, spectral calculus, etc.
Some of these basic concepts are outlined for the convenience of the reader.

Abbreviating we write IV for initial value(s), IVP for initial value problem(s), BV for boundary value(s), PDE for partial differential equation(s), and ONB for orthonormal basis. Moreover, the natural numbers are without zero, namely $\dN=\{1,2,3,4,5,...\}$.

%
%
\section{Wave-Like Equations in Hilbert Space}
\label{s1-general-wave}
Let $\cH$ be a separable real or complex Hilbert space with inner product $\skal..$, being anti-linear in the first and linear in
the second variable in the complex case, and with associated norm $\norm.=\sqrt{\skal..}$.
(Note: In some purely mathematical texts the inner product is taken linear in the first factor, but linearity in the second factor is general standard in mathematical physics. Also the notion $\skal..$ for the scalar product is common in mathematical physics, but in mathematics one also finds $(.,.)$ or $(.;.)$.)

\subsection{Preliminaries on Selfadjoint Operators, Spectral Calculus}
All operators used are linear, so we will not mention this anymore.
For readers who are not so familiar with the operator concept in Hilbert space theory, let us mention some basics.

Discontinuity of an operator $A$ (with respect to the $\norm.$--topology on $\cH$) is equivalent to its unboundedness. For such an unbounded operator $A$ its domain of definition $\dom(A)$ cannot be the whole Hilbert space $\cH$. Instead of that, an unbounded $A$ 
is defined on a (norm--) dense domain of definition $\dom(A)\subset\cH$, only,
making necessary a particularly careful mathematical treatment.
The differential operators acting in $\mathrm{L}^2$--Hilbert spaces,
which we will deal with in the next sections, all are unbounded.

For a possibly densely defined operator $A$ with domain $\dom(A)\subseteq\cH$, we repeat the following
notions common in Hilbert space theory:
\begin{enumerate}
\renewcommand{\labelenumi}{(\alph{enumi})}
\renewcommand{\theenumi}{(\alph{enumi})}
\item
$A$ is called \emph{positive}, if  $\skal{\xi}{A\xi}\geq0$ for all $\xi\in\dom(A)$.
\item
The \emph{adjoint} $A^*$ of $A$ is defined by
\begin{equation}
\label{s2-def-adjoint}
\begin{aligned}
&\dom(A^*)
=
\set{\xi \in \cH}{\text{$\exists\,\eta_\xi\in\cH$ with $\skal{\eta_\xi}{\vp}= \skal{\xi}{A\vp}$ $\forall \vp\in \dom(A)$}},
\\
&A^*\xi
=
\eta_\xi\,,\qquad\forall \xi\in\dom(A^*)\,.
\end{aligned}
\end{equation}
$\eta_\xi$ is unique since $\dom(A)$ is dense, and so the adjoint $A^*$ only exists for a densely defined operator $A$.
\item
$A$ is  \emph{symmetric}, if $A\xi=A^*\xi$ for all $\xi\in\dom(A)\subseteq\dom(A^*)$,
denoted by $A\subseteq A^*$, that is, if $\skal{\xi}{A\vp}=\skal{A\xi}{\vp}$ for all $\xi,\vp\in\dom(A)$.
\item
$A$ is called \emph{selfadjoint}, if $A=A^*$, implying symmetry but with the same domains of definition $\dom(A)=\dom(A^*)$.
\end{enumerate}

Selfadjointness -- and not only symmetry -- of an unbounded operator $A$ is an important property, since
only selfadjoint operators enable spectral calculus, e.g., \cite{Weidmann} Chapter\,8, \cite{ReedSimon} Vol.\,I Chapter\,VIII, for an overview \cite{HoneggerRiekers15} Section\,43.3.

Because we need the \textbf{spectral calculus} in Theorem~\ref{theo-1} below, let us repeat its essentiality in some detail.
Given a selfadjoint operator $A$ acting in $\cH$.
Spectral calculus is the basis for defining to every ordinary function $f$ an operator $f(A)$ acting in $\cH$.
Precisely this means, any real- or complex-valued function
\begin{equation*}
f:\sigma(A) \to \dR\, \text{(or $\dC$)}\,,\quad
y\mapsto f(y)
\end{equation*}
being defined on the spectrum $\sigma(A)\subseteq\dR$ of $A$, gives rise to an operator $f(A)$ acting in $\cH$.
$f(A)$ is selfadjoint, if and only if the ordinary function $f$ is real-valued. $f(A)$ is a bounded, thus a continuous operator, if $f$ is a bounded function. $f(A)$ is a positive operator on $\cH$, if $f$ has values only in the positives $[0,\infty)$.

Note that $A$ is positive if and only if the spectrum $\sigma(A)\subseteq [0,\infty)$.
Therefore, we will work with functions $f$ defined on $[0,\infty)$ when considering positive, selfadjoint operators $A$ as in the next subsection.

%

%
%
\subsection{Arbitrary Positive, Selfadjoint Operator $A$}
Let $A$  be a positive, selfadjoint operator acting in $\cH$. When $A$ is unbounded, then 
its domain of definition $\dom(A)$ has to be a proper dense subspace of $\cH$.
\begin{theo}[Wave-Like IVP]
\label{theo-1}
Consider the following wave-like IVP for the positive, selfadjoint operator $A$ in the Hilbert space $\cH$,
\begin{alignat}{2}
\label{eq:WEq-1}
&\text{wave-like differential equation}                     &\qquad     & \frac{d^2u(t)}{dt^2}=-Au(t)\,,  \quad t \in\dR\,,
\\   \notag
&\text{IV (at $t=0$)}                        &    & \left.u(t)\right|_{t=0}=u_0 \in\cH\,,
\\   \notag
&\text{IV (at $t=0$)}                        &    & \left.\frac{du(t)}{dt}\right|_{t=0}  = \dot{u}_0\in\cH\,,
\end{alignat}
with given Hilbert space vectors $u_0,\: \dot{u}_0\in\cH$.
Equation \eqref{eq:WEq-1} is  the short notion of the differential equation, it is mathematically rigourously formulated in the weak sense as
\begin{equation}
\label{eq:WEq-1-weak}
\frac{d^2}{dt^2}\skal{\eta}{u(t)}
=
-\skal{A\eta}{u(t)}\,,
\quad \forall \eta\in\dom(A)\,.
\end{equation}

Then the unique solution trajectory of the wave-like IVP is given by
\begin{equation}
\label{eq:WEq-sol-1}
u(t)=
\cos(t\sqrt{\smash[b]A})u_0  +
\frac{\sin(t\sqrt{\smash[b]A})}{\sqrt{\smash[b]A}}\dot{u}_0\,,
\quad\forall t\in\dR\,.
\end{equation}
Furthermore, the solution trajectory  $\dR\ni t\mapsto u(t)\in\cH$ is continuous with respect to the norm $\norm.$ of the Hilbert space $\cH$.
\end{theo}
\begin{skproof}
That \eqref{eq:WEq-sol-1} is indeed a solution of the IVP, is immediately verified with help of the spectral calculus.
For uniqueness see \cite{Leis86} Chapter 3, or \cite{Wloka82}.
\end{skproof}

Remark, as mentioned in the previous subsection, for each $t\in\dR$ the two selfadjoint operators $\cos(t\sqrt{\smash[b]A})$
and $\frac{\sin(t\sqrt{\smash[b]A})}{\sqrt{\smash[b]A}}$ arise by spectral calculus from the
ordinary continuous, bounded, real-valued functions of a single variable,
\begin{equation}
\label{s2-eq-sin(ty)/y}
[0,\infty)\ni y\mapsto \cos(t\sqrt{\smash[b]{y}})\,,
\qquad\quad
[0,\infty)\ni y\mapsto
\begin{cases}
t\,,                        & \text{if $y=0$}\,, \\
\frac{\sin(t\sqrt{\smash[b]{y}})}{\sqrt{\smash[b]{y}}}\,, & \text{if $y>0$}\,.
\end{cases}
\end{equation}
Regardless of whether $A$ is bounded or unbounded, both selfadjoint operators $\cos(t\sqrt{\smash[b]A})$
and $\frac{\sin(t\sqrt{\smash[b]A})}{\sqrt{\smash[b]A}}$ are bounded, and thus defined everywhere in $\cH$.
Therefore, the solution formula \eqref{eq:WEq-sol-1} is indeed valid for all IV Hilbert space vectors  $u_0,\: \dot{u}_0\in\cH$.
\begin{coro}
\label{coro-1}
If $\cH$ is a complex Hilbert space, then  the solution $t\mapsto u(t)$ from \eqref{eq:WEq-sol-1} is related to
the strongly continuous unitary group $\e^{it\sqrt{\smash[b]A}}$ in the following sense,
\begin{equation*}
u(t)=\e^{it\sqrt{\smash[b]A}}\,u_0\;\:\forall t\in\dR\,,\qquad\text{ if and only if }\qquad \dot{u}_0=i\sqrt{\smash[b]A}\,u_0\;\:\text{(at $t=0$)}\,.
\end{equation*}
\end{coro}
\subsection{Operator $A$ with Purely Discrete Spectrum}
\label{s1-discrete}
Let us suppose that the positive, selfadjoint operator $A$ in Theorem~\ref{theo-1} possesses a pure point (= purely discrete) spectrum, $\sigma(A)=\sigma_p(A)\subset[0,\infty)$, meaning a pure eigenspectrum: There exists an orthonormal basis
(ONB) of $\cH$, which consists of the \emph{normalized} eigenvectors $\psi_n$, $n\in\dN$, of our operator $A$, corresponding to the eigenvalues (= discrete spectral points) $a_n\geq0$, $n\in\dN$, in terms of the eigenequation
\begin{equation}
\label{eq:WEq-2-pps}
A\psi_n=a_n\psi_n\,,\quad \forall n\in\dN\,.
\end{equation}
(Note, $\cH$ is supposed to be separable, so the ONB is countable.)
Thus, for every ordinary function $f:[0,\infty)\ra\dC, y\mapsto f(y)$ it follows by spectral calculus that 
\begin{equation}
\label{eq:WEq-2-pps-f}
f(A)\psi_n=f(a_n)\psi_n\,,\quad \forall n\in\dN\,. 
\end{equation}
\begin{coro}
\label{coro-1-discrete}
With the purely discrete spectrum \eqref{eq:WEq-2-pps} of the positive, selfadjoint operator $A$, 
the solution trajectory $t\mapsto u(t)\in\cH$ of \eqref{eq:WEq-sol-1} rewrites as
\begin{equation}
\label{eq:WEq-sol-2}
u(t)
=
\sum_{n=1}^\infty
\Bigl(
\underbrace{
\cos(t\sqrt{\smash[b]{a_n}})\skal{\psi_n}{u_0}  +
\frac{\sin(t\sqrt{\smash[b]{a_n}})}{\sqrt{\smash[b]{a_n}}}\skal{\psi_n}{\dot{u}_0}
}_{\mbox{$=\,\skal{\psi_n}{u(t)}$}}
\Bigr)
\psi_n\,,
\quad\forall t\in\dR\,.
\end{equation}
This is the decomposition of the wave-like solution $u(t)$ into the eigenvectors $\psi_n$ of the positive, selfadjoint operator $A$,
also denoted as eigenvector expansion of $u(t)$.
\end{coro}
\begin{proof}
First remark the spectral properties by equation \eqref{eq:WEq-2-pps-f},
\begin{equation}
\label{eq:WEq-sol-2-proof}
\cos(t\sqrt{\smash[b]A})\psi_n=\cos(t\sqrt{\smash[b]{a_n}})\psi_n\,,
\quad
\textstyle\frac{\sin(t\sqrt{\smash[b]A})}{\sqrt{\smash[b]A}}\psi_n=\frac{\sin(t\sqrt{\smash[b]a_n})}{\sqrt{\smash[b]a_n}}\psi_n\,,
\quad
\forall n\in\dN\,.  
\end{equation}
Since the normalized eigenvectors $\psi_n$, $n\in\dN$ constitute an ONB of $\cH$, we may decompose $u(t)$ of \eqref{eq:WEq-sol-1} according to the
spectral projections
\begin{align*}
u(t)
&=
\sum_{n=1}^\infty \skal{\psi_n}{u(t)}\psi_n
\\
&=
\sum_{n=1}^\infty
\skal{\psi_n}{\cos(t\sqrt{\smash[b]{A}})u_0 + {\textstyle\frac{\sin(t\sqrt{\smash[b]{A}})}{\sqrt{\smash[b]{A}}}}\dot{u}_0}
\psi_n
\\
&=
\sum_{n=1}^\infty
\Bigl(
\skal{\psi_n}{\cos(t\sqrt{\smash[b]{A}})u_0}  +
\skal{\psi_n}{{\textstyle\frac{\sin(t\sqrt{\smash[b]{A}})}{\sqrt{\smash[b]{A}}}}\dot{u}_0}
\Bigr)
\psi_n
\\
&\stackrel{\star}=
\sum_{n=1}^\infty
\Bigl(
\skal{\cos(t\sqrt{\smash[b]{A}})\psi_n}{u_0}  +
\skal{{\textstyle\frac{\sin(t\sqrt{\smash[b]{A}})}{\sqrt{\smash[b]{A}}}}\psi_n}{\dot{u}_0}
\Bigr)
\psi_n
\\
&\stackrel{\eqref{eq:WEq-sol-2-proof}}=
\sum_{n=1}^\infty
\Bigl(
\cos(t\sqrt{\smash[b]{a_n}})\skal{\psi_n}{u_0}  +
{\textstyle\frac{\sin(t\sqrt{\smash[b]{a_n}})}{\sqrt{\smash[b]{a_n}}}}\skal{\psi_n}{\dot{u}_0}
\Bigr)
\psi_n\,,
\end{align*}
where at the equality sign $\stackrel{\star}=$ with the star we used the selfadjointness of the two bounded operators $\cos(t\sqrt{\smash[b]{A}})$ and
${\textstyle\frac{\sin(t\sqrt{\smash[b]{A}})}{\sqrt{\smash[b]{A}}}}$.
\end{proof}

Note, if the spectrum of $A$ contains continuous parts (possibly whole intervals), then the solution cannot be written in such a simple way.

\begin{rema}[Modification by a Physical or Material Constant]
\label{s3-substitute}
In the next sections we consider such positive, selfadjoint operators $A$ for physical, engineering, or technical applications.
In general there $A$ is modified to $\varsigma^2A$ with some physical or material constant $\varsigma>0$ 
in the wave-like differential equation \eqref{eq:WEq-1},
\begin{equation*}
\text{modified wave-like differential equation}\qquad\frac{d^2u(t)}{dt^2}=-\varsigma^2Au(t)\,,  \quad t \in\dR\,.
\end{equation*}
Then $\sqrt{\smash[b]A}$ in our solution formula \eqref{eq:WEq-sol-1} has to be replaced by $\varsigma\sqrt{\smash[b]A}$, 
implying that $\sqrt{\smash[b]{a_n}}$ is substituted by $\varsigma\sqrt{\smash[b]{a_n}}$ in the solution \eqref{eq:WEq-sol-2} for eigenvector expansion.
\end{rema}
%
%

%
%
\section{About Applications in $\mathrm{L}^2$--Hilbert Spaces}
\label{s2-Weq-in-L2}
Let $\cH=\mathrm{L}^2(\Lam)$ be the Hilbert space of $\dR$-- or $\dC$--valued,
Lebesgue square integrable functions defined on the spatial region $\Lam\subseteq\dR^r$ in $r\in\dN$ real dimensions,
with the standard inner product ($\ovl{\xi(x)}$ complex conjugate to $\xi(x)$) and Hilbert space norm,
\begin{equation*}
\skal{\xi}{\eta}=\int_{\Lam}\ovl{\xi(x)}\eta(x)\, d^r x\,,
\quad
\norm{\xi}^2=\skal{\xi}{\xi}=\int_{\Lam}\betr{\xi(x)}^2 d^r x\,,
\qquad \forall \xi,\eta\in\mathrm{L}^2(\Lam)\,.
\end{equation*}
$\Lam$ is chosen as an \emph{open} and \emph{connected} subset of $\dR^r$, which usually is called a \emph{domain}
or a \emph{region}.
``Connected'' means ``path connected'', so that any pair of points in $\Lam$
may be connected via a continuous path within $\Lam$. The domain
$\Lam$ is called \emph{interior} if $\Lam$  is bounded,
and an \emph{exterior} region if its set complement $\dR^r\!\setminus\!\Lam$ is bounded.
$\bar{\Lam}$ denotes the topological closure of the domain $\Lam$, and so
$\partial\Lam=\bar{\Lam}\!\setminus\!\Lam$ is just the boundary of  $\Lam$.

The subsequent Subsections~\ref{ss3.1-wavelike} to~\ref{s3-bending-plate}  
are dedicated to special examples of wave-like equations formulated in $\mathrm{L}^2$--Hilbert space terminology.
Subsections~\ref{s2-spectral-properties} and~\ref{ss-2:regularity} summarize some general properties of related differential operators and their spectra.
\begin{proc}[Wave-Like Differential Operators in $\mathrm{L}^2$--Hilbert Space]
One starts from partial differentiation within a region $\Lam\subseteq\dR^r$ with respect to the $r$ spatial variables. 
The first step is to transform this spatial differentiation operation on $\Lam$ into a positive, selfadjoint operator $A$ acting in the Hilbert space $\mathrm{L}^2(\Lam)$, a process, which in general requires much mathematical effort.
Now $A$ can be taken for the wave-like differential equation in Section \ref{s1-general-wave}, especially Theorem \ref{theo-1}.

The exact mathematical definition of such a positive, selfadjoint differential operator $A$ acting in $\mathrm{L}^2(\Lam)$ is often done in terms
of a closed, positive sesquilinear form by Friedrichs extension \cite{Kato84,Weidmann}, which necessarily has to include the considered boundary condition.
That means, the chosen BV are intrinsically involved into the positive, selfadjoint differential operator $A$,
and are not an extra condition. Possibly the considered BV require some kind of smoothness for the boundary $\partial\Lam$ of the region $\Lam$,
e.g.\ segment property, or uniform cone property, or piece-wise $\operatorname{C}^k$--smoothness, etc.

Finally, the spectral properties of the positive, selfadjoint differential operator $A$ have to be worked out. In general they turn out to be as mentioned in the Subsections~\ref{s2-spectral-properties} and~\ref{ss-2:regularity}, e.g.\ \cite{Wloka82,Leis86,DautrayLions90,ReedSimon,Kato84}.
\end{proc}
In Section~\ref{s3-bending-vibs} this procedure is  performed for differential operators of $4$-th order
describing the tranversal bending vibrations of a slender beam in one spatial dimension.

\subsection{Primary Wave PDE in a Spatial Region $\Lam$}
\label{ss3.1-wavelike}
Consider the Laplace operator on an arbitrary spatial region $\Lam\subseteq\dR^r$,
\begin{equation*}
-\Delta = -\Bigl(\partial_1^2+\partial_2^2+\ldots+\partial_r^2\Bigr)\,,
\end{equation*}
with partial differentiation $\partial_j=\frac{\partial}{\partial x_j}$ in the $j$-th variable $x_j$, where $j=1,2,\ldots, r$.

It is well known that the Laplacian $-\Delta$ leads indeed to a positive and selfadjoint operator $A=-\Delta$
acting on $\mathrm{L}^2(\Lam)$
for each of the classical \emph{homogeneous} boundary conditions such as Dirichlet, or Neumann, or mixed.
Even when incorporating an anisotropic, inhomogeneous medium into $\Lam$, in which the waves propagate, with associated generalized Laplacian, then one may also show positivity and selfadjointness in many cases, e.g.\ \cite{Wloka82,Leis86,DautrayLions90}.

Here the differential equation \eqref{eq:WEq-1} describes the classical propagating wave in $\Lam$, which
satisfies the BV for which the positive, selfadjoint Laplacian $A=-\Delta$ is defined.
The solution trajectory $t\mapsto u(t)\in\mathrm{L}^2(\Lam)$ consists of ($\mathrm{L}^2$--classes of) functions
\begin{equation}
\label{eq:WEq-3}
u(t)(x_1,x_2,\ldots,x_r)= u(x_1,x_2,\ldots,x_r,t)\,,\quad \forall (x_1,x_2,\ldots,x_r)\in\Lam\,,\quad  t\in\dR\,.
\end{equation}
In terms of two times continuously differentiable  functions, the differential equation \eqref{eq:WEq-1}
is rewritten as the well known ordinary wave PDE, namely,
\begin{equation}
\label{eq:WEq-3b}
\partial_t^2 u(x_1,\ldots,x_r,t)
=
c^2\Bigl(\underbrace{\partial_1^2+\ldots+\partial_r^2}_{=\,\Delta}\Bigr) u(x_1,\ldots,x_r,t)
=-c^2(\underbrace{-\Delta}_{A\,\geq\,0}) u(x_1,\ldots,x_r,t)
\end{equation}
with wave velocity $c>0$. Here $\partial_t=\frac{\partial}{\partial t}$ indicates the partial differentiation with respect to the time variable $t\in\dR$.

\subsection{Maxwell Radiation and Wave Equations in Electromagnetism}
\label{ss3.2-wavelike-Maxwell}
For the details to the present subsection, the reader is referred to \cite{HoneggerRiekers15} Chapters 4 and 44.
We only give a short overview concerning wave equations for  electromagnetic radiation.

We consider the Maxwell equations in vacuum in the spatial region  $\Lam\subseteq\dR^3$. We use the real Hilbert space $\mathrm{L}^2(\Lam,\dR^6)$ for $\dR^6$--valued functions on $\Lam$ (vector-valued functions with 6 components).
By $\bE_t$ it is denoted the electric field and by $\bB_t$ the magnetic field (three components for each),
depending on time $t\in\dR$. Let us first describe their dynamical behaviour directly in terms of the Maxwell equations.

For simplicity we assume no current and no charge distribution in $\Lam$, implying divergence freeness according to the Maxwell
equations,
\begin{equation}
\label{ss-Maxwell-eq-0-divfree}
\div\bE_t = 0\,,\quad\div_0\bB_t = 0\,,\quad\forall t\in\dR\,.
\end{equation}
The remaining two Maxwell equations are of dynamical nature and are summarized in matrix notation by
\begin{equation}
\label{ss-Maxwell-eq-1}
\frac{d}{dt}
\underbrace{
\begin{pmatrix}
\bE_t\\ \bB_t
\end{pmatrix}
}_{\mbox{$=u(t)$}}
=
\underbrace{
\begin{pmatrix}
0 &\curl \\ -\curl_0 & 0
\end{pmatrix}
}_{\mbox{$=\dA$}}
\underbrace{
\begin{pmatrix}
\bE_t\\ \bB_t
\end{pmatrix}
}_{\mbox{$=u(t)$}}
\,,\quad\forall t\in\dR\,.
\end{equation}
The dielectric constant $\eps_0$ and
the magnetic permeability $\mu_0$ are set to $1$ for the convenience of the reader.

The walls of $\Lam$, that is the boundary $\partial\Lam$,  are supposed to consist of a perfect conductor material.
This leads to the well known boundary conditions
\begin{equation}
\label{eq:vt2:BC-pc}
\bE_t\times n|_{\partial\Lam}=0\,,
\quad
\bB_t\cdot n|_{\partial\Lam}=0\,,
\quad
\forall t\in\dR\,,
\end{equation}
where $n$ denotes the outer normal vector at the boundary points.
The two divergence operators, $\div_0$ and $\div$, as well as the two curl (rotation) operators, $\curl_0$ and $\curl$,
are adapted to these perfect conductor boundary conditions, with minimal and maximal Sobolev domains of definition, respectively.

The Maxwell operator $\dA$ is anti-selfadjoint, meaning $\dA^*=-\dA$ (that is, in the complexified Hilbert space $i\dA$ is selfadjoint), 
since $\curl^*=\curl_0$ and
$\curl_0^*=\curl$ for the adjoints.
Consequently, $\ex{t\dA}$, $t\in\dR$, constitutes
a strongly continuous orthogonal group in $\mathrm{L}^2(\Lam,\dR^6)$.
With given IV
\begin{equation*}
\left.u(t)\right|_{t=0}=u_0 =
\left(\begin{smallmatrix}
\bE_0\\ \bB_0
\end{smallmatrix}
\right)\,,
\end{equation*}
the unique solution of the IVP \eqref{ss-Maxwell-eq-1} turns out to be
\begin{equation}
\label{s2-maxwell-sol}
u(t)=
\ex{t\dA}u_0\,,
\quad\forall t\in\dR\,.
\end{equation}
Formula (\ref{s2-maxwell-sol}) describes the freely evolving electromagnetic field in the spatial region $\Lam$, namely the electromagnetic radiation,
in which intense coupling of the electric and magnetic fields takes place due to the component mixing by the \emph{nondiagonal} Maxwell matrix operator $\dA$ in equation~\eqref{ss-Maxwell-eq-1}.

It is well known that the electric and the magnetic components can be decoupled.
Taking the second time derivative in \eqref{ss-Maxwell-eq-1}, we arrive at the wave-like equation
\begin{equation}
\label{ss-Maxwell-eq-4}
\frac{d^2 u(t)}{dt^2}=\dA^2 u(t)=-\underbrace{\dA^*\dA}_{\geq\,0} u(t)\,,
\end{equation}
where we have inserted $\dA^*=-\dA$. It follows that
\begin{equation*}
\dA^*\dA=-\dA^2
=
-
\begin{pmatrix}
0 &\curl \\ -\curl_0 & 0
\end{pmatrix}
\begin{pmatrix}
0 &\curl \\ -\curl_0 & 0
\end{pmatrix}
=
\begin{pmatrix}
\curl \curl_0 & 0
\\
0 & \curl_0 \curl
\end{pmatrix}
\end{equation*}
is a \emph{diagonal} matrix operator in the electromagnetic field Hilbert space $\mathrm{L}^2(\Lam,\dR^6)$, which thus decouples the electric and magnetic fields. $\dA^*\dA$ is a positive and selfadjoint operator, and so are both double curl operators
$\curl \curl_0$ and $\curl_0\curl$.

The two positive, selfadjoint curlcurl--operators, $\curl \curl_0$ and $\curl_0\curl$, agree
with two \emph{different} Laplace operators, denoted by $-\Delta_E$ and $-\Delta_B$, corresponding 
to the perfect conductor boundary conditions \eqref{eq:vt2:BC-pc},
but are not covered by the mentioned classical BV cases in the previous subsection.
Decoupling ensures that we get two separate wave equations,
one for the electric field and another for the magnetic field,
\begin{equation*}
\frac{d^2 \bE_t}{dt^2}=-\overbrace{\curl \curl_0}^{=\,-\Delta_E\,\geq\,0} \bE_t\,,\qquad\quad
\frac{d^2 \bB_t}{dt^2}=-\overbrace{\curl_0 \curl}^{=\,-\Delta_B\,\geq\,0} \bB_t\,,
\end{equation*}
both living now in the Hilbert space $\mathrm{L}^2(\Lam,\dR^3)$ with three components, only.

In direct analogy to Corollary~\ref{coro-1}, the solutions of both wave equations agree with the original solution \eqref{s2-maxwell-sol}
of the dynamical Maxwell equations \eqref{ss-Maxwell-eq-1} exclusively, when the subsequent specific correlation of the IV
at the initial time point $t=0$ is fulfilled, 
$$
\underbrace{
\begin{pmatrix}
\dot{\bE}_0\\ \dot{\bB}_0
\end{pmatrix}
}_{\mbox{$=\dot{u}_0$}}
=
\left. \frac{d u(t)}{dt}\right|_{t=0} 
\!=
\left. \frac{d \ex{t\dA}u_0}{dt}\right|_{t=0} 
\!=
\underbrace{
\begin{pmatrix}
0 &\curl \\ -\curl_0 & 0
\end{pmatrix}
}_{\mbox{$=\dA$}}
\underbrace{
\begin{pmatrix}
\bE_0\\ \bB_0
\end{pmatrix}
}_{\mbox{$=u_0$}}
=
\begin{pmatrix}
\curl \bB_0\\ -\curl_0\bE_0
\end{pmatrix}
.
$$

\subsection{Free Bending Vibrations of a Plate}
\label{s3-bending-plate}
The plate is concentrated in the spatial region $\Lam\subseteq\dR^r$, and bends transversally into an additional spatial dimension.
For an isotropic, homogeneous plate the operator $A$ of Section \ref{s1-general-wave} is given e.g.\ with $r=2$ by
\begin{equation*}
-\Delta^2 = -\bigl(\partial_1^2+\partial_2^2\bigr)^2
  = -\Bigl(\partial_1^4+2\partial_1^2\partial_2^2+\partial_2^4\Bigr)\,,
\end{equation*}
up to some material constant. One may also incorporate anisotropy and inhomogeneity for the plate.
In the literature one finds some homogeneous BV, for which such an operator $A$
turns out to be positive and selfadjoint, see e.g.\ \cite{Leis86}.

The literature, however, does not cover the case $r=1$ with the diverse BV from engineering statics and technical mechanics,
which we will investigate in great mathematical detail in Section~\ref{s3-bending-vibs}.

\subsection{On Spectral Properties of the Differential Operators $A$}
\label{s2-spectral-properties}
Positive, selfadjoint differential operators $A$ of the above types possess (in general) the following spectral properties,
\begin{alignat*}{2}
&\text{pure point (= purely discrete) spectrum $\sigma(A)=\sigma_{p}(A)\subset[0,\infty)$} & & \text{for interior $\Lam$},
\\
&\text{absolutely continuous spectrum $\sigma(A)=\sigma_{ac}(A)=[0,\infty)$} &\quad& \text{for exterior $\Lam$},
\end{alignat*}
with some mild assumptions about the smoothness of the boundary $\partial\Lambda$ of the region $\Lam$.
In addition, for interior $\Lam$ each eigenspace is finite dimensional and the eigenvalues $a_n$, $n\in\dN$,
from \eqref{eq:WEq-2-pps} may be arranged increasingly and converge to infinity,
\begin{equation*}
0\leq a_1\leq a_2\leq a_3\leq a_4\leq a_5 \leq .....
\,,\qquad\quad
\lim_{n\ra\infty} a_n=\infty\,.
\end{equation*}
Recall from Subsection \ref{s1-discrete}, a purely discrete spectrum is just a pure eigenspectrum of $A$, namely:
$A\psi_n=a_n\psi_n$ for all $n\in\dN$, where
the $a_n\geq0$ are the eigenvalues with
corresponding normalized eigenvectors $\psi_n$. In addition, the set of eigenvectors 
$\{\psi_n\mid n\in\dN\}$ constitutes an ONB of the Hilbert space.

There is, however, an exception for the two curlcurl--Laplacians in electromagnetism, $-\Delta_E=\curl \curl_0$ and $-\Delta_B=\curl_0\curl$,
respectively. The kernels of these both operators
(= eigenspaces to eigenvalue zero) are infinite dimensional. These kernels consist of suitable divergence-free fields in accordance with \eqref{ss-Maxwell-eq-0-divfree}.
For all other eigenvalues, that are the strictly positive eigenvalues $a_n>0$, the preceding statements are valid, including
finite multiplicity and convergence to infinity.

In general the existence of a discrete spectrum for interior $\Lam$ is  proven with the help of compact embeddings of related Sobolev spaces into $\mathrm{L}^2(\Lam)$, e.g.\ \cite{Wloka82,DautrayLions90}.
Smoothness properties of the associated eigenvectorfunctions are given by regularity arguments,
cf.\ the next Subsection~\ref{ss-2:regularity}.
We also use that argument with a compact embedding in section \ref{s6-proofs} for proving
the spectral properties of diverse positive, selfadjoint differential operators of 4-th order in Theorem \ref{s3-theo-pos-sa-extension-2}
for the Euler--Bernoulli bending vibrations of a slender beam in the bounded (= interior) interval $(0,\ell)$.

Exterior domains $\Lam$ with absolutely continuous spectra are commonly used in scattering theory,
e.g.\ \cite{AmreinJauchSinha77,ReedSimon}, and many more.

\subsection{On Regularity of Eigenvectors and Solutions}
\label{ss-2:regularity}
In the cited and further literature one may find many results concerning  regularity.
Regularity statements are of the following kind:
\begin{enumerate}
\renewcommand{\labelenumi}{(\alph{enumi})}
\renewcommand{\theenumi}{(\alph{enumi})}
\item
Smoothness of the boundary $\partial\Lam$
implies smoothness of the eigenvectors $\psi_n=\psi_n(x_1,\ldots,x_r)$, interior $\Lam$ supposed.
\item
Regularity of a Hilbert space solution function $u(t)$ of \eqref{eq:WEq-3} implies smoothness of the solution function $u(x_1,\ldots,x_r,t)$,
depending on the smoothness degrees of the boundary $\partial\Lam$ and
of the two IV functions $u_0,\, \dot{u}_0\in\cH=\mathrm{L}^2(\Lam)$.
\end{enumerate}
For the detailed definition of smoothness, i.e.\ continuity or differentiability properties,
see e.g.\ \cite{Wloka82,Leis86,DautrayLions90,HoneggerRiekers15}.

\begin{summ}[The Importance of Regularity]
\label{s3-summ=regularity}
In general, every $\mathrm{L}^2$--Hilbert space element
$\upsilon\in\mathrm{L}^2(\Lam)$ is a class consisting of functions represented by $\upsilon$ with values either in $\dR$ or in $\dC$,
\begin{equation}
\label{eq:WEq-10}
\Lam\ni(x_1,x_2,\ldots,x_r)\;\mapsto\;
\upsilon(x_1,x_2,\ldots,x_r)\in\dR\text{\ \ or $\in\dC$}.
\end{equation}
All functions represented by $\upsilon$  agree \emph{almost everywhere} in $\Lam$, only (with respect to the common Lebesgue measure $d^r x$ on $\Lam$). 
Therefore, a  point evaluation or a conventional partial differentiation does make no sense.

That is the reason, why a Hilbert space formulation of differential equations
is always a generalization, which has to be formulated weakly like in equation \eqref{eq:WEq-1-weak}.

When in addition the class $\upsilon$ contains an element $u$ -- a function -- with some smoothness properties (= regularity),
then that representant $u$ of the class $\upsilon$ allows for a point evaluation or ordinary partial differentiations defined by differential limits.
And then the $\mathrm{L}^2$--solution trajectory $t\mapsto u(t)$ as in equation \eqref{eq:WEq-3}
fulfilling the $\mathrm{L}^2$--Hilbert space wave-like equation, solves the ordinary wave-like PDE
in the classical or traditional sense with ordinary partial derivatives in terms of conventional differential limits.
\end{summ}

\vspace{3ex}
So far in the present section we have presented a short overview of results from functional analysis.
But, as mentioned already in Section~\ref{s0-introduction}, for a \emph{general} interior $\Lam$
it is  not possible to compute analytically the eigenvectors (= eigenfunctions)
of a positive, selfadjoint differential operator $A$ acting in $\mathrm{L}^2(\Lam)$.
Nevertheless Hilbert space theory is able
to prove the existence of an eigenspectrum, several regularity results, and  more.
Concrete calculations of eigenvalues and eigenvectorfunctions are in general possible, only,
if $\Lam$ has certain geometric properties such as symmetry. Parallelepiped, ball, or circular disc, are standard textbook examples,
e.g.\ \cite{DautrayLions90}, and references therein.

The case $r=1$ of a single spatial dimension is obviously such a special case.
There an interior domain $\Lam$ is always an open bounded interval, 
e.g.\ $\Lam=(0,\ell)$ for $\ell>0$ as taken in the subsequent sections.
The interval $(0,\ell)$ possesses a ``completely smooth boundary'', namely its edge points $x=0$ and $x=\ell$. 
So by regularity arguments, eigenfunctions and solution trajectories should be smooth,
provided some smoothness degrees of the IV $u_0$ and $\dot{u}_0$.

%
%
\section{Differential Operators for the Interval $(0,\ell)$}
\label{s3-sobolev+diffOps}

In the interior open interval $\Lam=(0,\ell)$ with  boundary (= edge) points $x=0$ and $x=\ell$ 
is placed in the next section a slender beam.
Before we arrive at the wave-like Euler--Bernoulli differential equation for bending vibrations of that beam, as preparation it is first necessary
to introduce several differential operators of first and second order.

Let us abbreviate $\mathrm{L}^2 := \mathrm{L}^2((0,\ell))$ for 
the complex Hilbert space of $\dC$-valued, square integrable functions $\xi:(0,\ell)\ra\dC,\, x\mapsto\xi(x)$
with inner product and norm
\begin{equation*}
\skal{\xi}{\eta}=\int^\ell_0\ovl{\xi(x)}\eta(x)\, dx\,,\quad
\norm{\xi}^2=\skal{\xi}{\xi}=\int^\ell_0\betr{\xi(x)}^2 dx\,,
\quad \forall \xi,\eta\in\mathrm{L}^2 = \mathrm{L}^2((0,\ell))\,.
\end{equation*}

\subsection{Sobolev Spaces for the Interior Interval $(0,\ell)$}
\label{ss-3:Sobolev-space}
For the mathematical description of differential operators it is inevitable to work with Sobolev spaces.
For completeness we state here some notions and properties we need subsequently.

By $\operatorname{C}_c^\infty(I)$
we denote the set of infinitely often continuously differentiable functions $\xi:I\ra\dC$
for the open interval $I\subseteq\dR$
with compact support within $I$, the  standard testfunction space in distribution theory for $I$.
The elements of
$\operatorname{C}_c^\infty(I)|_J$
are the restrictions $\xi|_J$ of
$\xi\in\operatorname{C}_c^\infty(I)$ to the open subinterval $J\subseteq I$.

Let $\xi:\dR\ra\dC$ be an $s$--times continuously differentiable function and $\vp$ a testfunction on $(0,\ell)$, that is $\vp\in \operatorname{C}_c^\infty((0,\ell))$.
When integrating $s$--times partially, no boundary terms occur, since the testfunction $\vp$ has compact support
in $(0,\ell)$ and hence vanishing boundary values $\vp^{(k)}(0)=0=\vp^{(k)}(\ell)$ for all derivatives,
\begin{equation*}
\skal{\xi^{(s)}}{\vp}
=
\int_0^\ell \overline{\xi^{(s)}(x)}\vp(x)\,dx
=
(-1)^s\int_0^\ell \overline{\xi(x)}\vp^{(s)}(x)\,dx
=
(-1)^s\skal{\xi}{\vp^{(s)}}
\,.
\end{equation*}
This is the guiding line for the introduction of the following generalized concept of differentiability.
\begin{defi}[Square Integrable Distributional Differentiability]
\label{s3-defi-sidd}
Suppose for a $\xi\in \mathrm{L}^2$ the existence of a vector $\xi^{(s)}\in \mathrm{L}^2$ such that
\begin{equation}\label{eq:derivative}
\skal{\xi^{(s)}}{\vp}
=
(-1)^s\skal{\xi}{\vp^{(s)}}\,,
\quad\forall \vp\in\operatorname{C}_c^\infty((0,\ell))\,.
\end{equation}
Then $\xi^{(s)}\in \mathrm{L}^2$ is called the square integrable $s$-th derivative of $\xi\in\mathrm{L}^2$ (exactly, differentiability in the distributional sense).
\end{defi}
Note that, provided existence, the vector $\xi^{(s)}\in \mathrm{L}^2$ is unique because the testfunction space $\operatorname{C}_c^\infty((0,\ell))$
is $\norm.$--dense in $\mathrm{L}^2$.

As shown above, $\xi^{(s)}$ agrees with the conventional higher derivative, whenever $\xi$ is $s$--times continuously differentiable,
and so the distributional definition is an extension of ordinary differentiation using conventional differential limits.
\begin{defi}[Sobolev Spaces]
\label{s3-Spaces-defi}
For each $m\in\dN_0$ the $m$-th Sobolev space is defined as
\begin{equation*}
\mathrm{W}^m= \mathrm{W}^m((0,\ell)):=
\{\xi\in\mathrm{L}^2\mid \exists\,\xi^{(s)}\in\mathrm{L}^2\text{\ \ for\ \ }0\leq s\leq m\}\,.
\end{equation*}
It is equipped with the inner product
\begin{equation}
\label{eq:qbdiag:Sobolev-IP}
\skal{\xi}{\eta}_m:=
\sum_{s=0}^m \skal{\xi^{(s)}}{\eta^{(s)}}
\,,\qquad \forall \xi,\eta\in\mathrm{W}^m,
\end{equation}
with associated $m$-th Sobolev norm $\norm{\xi}_m=\sqrt{\skal{\xi}{\xi}_m}$.
\end{defi}
The index $m=0$ yields
$\skal{\xi}{\eta}_0=\skal{\xi}{\eta}$ and $\norm{\xi}_0=\norm{\xi}=\sqrt{\skal{\xi}{\xi}}$ being the conventional
scalar product and norm on $\mathrm{L}^2=\mathrm{W}^0$ with $\xi=\xi^{(0)}$.

We present some results from the literature, see e.g.\ \cite{Wloka82}, an overview is found in \cite{HoneggerRiekers15}\,Subsection\,44.1.2.
\begin{prop}[Properties]
\label{s3-prop-W-properties}
The following assertions are valid:
\begin{enumerate}
\renewcommand{\labelenumi}{(\alph{enumi})}
\renewcommand{\theenumi}{(\alph{enumi})}
\item
\label{s3-prop-W-properties-a}
$\mathrm{W}^m$ is a separable complex
Hilbert space for every $m\in\dN_0$ with respect to its Sobolev inner product $\skal{.}{.}_m$ and
$m$-th norm  $\norm{.}_m$ from \eqref{eq:qbdiag:Sobolev-IP}.
\item
\label{s3-prop-W-properties-b}
$\left.\operatorname{C}_c^\infty(\dR)\right|_{(0,\ell)}$ is $\norm._m$--dense
in the $m$-th Sobolev space $\mathrm{W}^m$ for each $m\in\dN_0$.
(This way $\mathrm{W}^m$ may be defined without distributional derivatives.)
\item
\label{s3-prop-W-properties-c}
If $m>k$,  then
$\mathrm{W}^m\subseteq \operatorname{C}^k([0,\ell])$,
the $k$--times continuously differentiable functions on the open interval $(0,\ell)$, for which each derivative
extends continuously to both boundary points $x=0$ and $x=\ell$.
\item
\label{s3-prop-W-properties-d}
Let $m>n$. Then the identical embedding
$\mathrm{W}^{m}\hookrightarrow \mathrm{W}^{n}$
is continuous, injective and a compact map.
Especially,  the identical embedding $\mathrm{W}^{1} \hookrightarrow \mathrm{L}^2$ is compact.
\end{enumerate}
\end{prop}

Note, here in one dimension, an absolutely continuous function $\xi:(0,\ell)\ra\dC$, $x\mapsto \xi(x)$
is differentiable almost everywhere in the sense of conventional differentiation,
and its  derivative $\xi'$ is integrable over all compact subintervals of $(0,\ell)$,
but not necessarily square integrable. This way the  first distributional derivative $\xi'$
is related to ordinary differentiation by differential limits, provided square integrability is assumed.

\subsection{Differential Operators of First Order for Diverse BV}
\label{ss-3:diff-op}
We define four different differential operators $\delta_{.\:.}\in\{\delta_{++},\delta_{+-},\delta_{-+},\delta_{--} \}$ 
of first order acting on $\mathrm{L}^2$ (differentiation in the distributional sense),
\begin{equation*}
\delta_{.\:.}\,\xi=\xi'\,,\qquad \forall \xi\in \dom(\delta_{.\:.})\subset \mathrm{L}^2\,.
\end{equation*}
As mentioned, unboundedness (= discontinuity) makes it impossible that such a differential operator may act on all Hilbert space vectors.
Therefore we define four different domains of definition leading to four different operators.
The following domains of definition $\dom(\delta_{.\:.})\subset\mathrm{L}^2$ are $\norm.$--dense
in $\mathrm{L}^2$, they are according to the subsequent construction indirectly adapted to
the following boundary conditions at the two boundary points $x=0$ and $x=\ell$ of the open interval $(0,\ell)$,
\begin{alignat*}{3}
&\delta_{++}           &\quad\;     & \dom(\delta_{++})=\norm._1\text{--closure of }\operatorname{C}_c^\infty((0,\ell)),
&\text{\ \ \  BV: }& \xi(0)=0\,,\;\; \xi(\ell)=0\,;
\\
&\delta_{+-}           &     & \dom(\delta_{+-})=\norm._1\text{--closure of }
\left.\operatorname{C}_c^\infty((0,\infty))\right|_{(0,\ell)},
&\text{\ \ \  BV: }& \xi(0)=0\,, \text{ no BV at $\ell$}\,;
\\
&\delta_{-+}           &     & \dom(\delta_{-+})=\norm._1\text{--closure of }
\left.\operatorname{C}_c^\infty((-\infty,\ell))\right|_{(0,\ell)},
&\text{\ \ \ BV: }& \text{no BV at $0$},\;\xi(\ell)=0\,;
\\
&\delta_{--}           &     & \dom(\delta_{--})=\norm._1\text{--closure of }
\left.\operatorname{C}_c^\infty(\dR)\right|_{(0,\ell)}=\mathrm{W}^1,
&\text{\ \ \  BV: }& \text{no BV at both $0$ and $\ell$}\,.
\end{alignat*}
The  minus or plus sign in the index means: For ``$+$'' the BV is fulfilled, and for ``$-$'' the BV is not fulfilled,
corresponding to the left or right boundary point, $x=0$ and $x=\ell$, respectively.
The closures of these spaces with respect to the first Sobolev norm $\norm._1$ are defined to be
the domains of definition $\dom(\delta_{.\:.})\subset\mathrm{L}^2$ of these operators $\delta_{.\:.}$ regarded as operators acting in $\mathrm{L}^2$.
Of course, by construction the domains $\dom(\delta_{.\:.})$ are $\norm._1$--closed subspaces of
the Sobolev Hilbert space $\mathrm{W}^1$.

From Proposition \ref{s3-prop-W-properties}\ref{s3-prop-W-properties-c} we know that $\mathrm{W}^1$ is a subspace of $ \operatorname{C}^0([0,\ell])$ (continuous functions on the closed interval $[0,\ell]$) and thus the above abstract definition
with $\norm._1$--closures allows for a direct boundary evaluation, in contrast to the previous indirect BV construction,
and it follows with Proposition \ref{s3-prop-W-properties}\ref{s3-prop-W-properties-b},
\begin{alignat*}{2}
&\delta_{++}  &\quad\quad & \dom(\delta_{++})=\set{\xi\in\mathrm{W}^1}{\xi(0)=0\,,\;\; \xi(\ell)=0}\,;
\\
&\delta_{+-}   &\quad\; & \dom(\delta_{+-})=\set{\xi\in\mathrm{W}^1}{\xi(0)=0\,, \text{ no BV at $\ell$}}\,;
\\
&\delta_{-+}   &\quad\; & \dom(\delta_{-+})=\set{\xi\in\mathrm{W}^1}{\text{no BV at $0$},\;\xi(\ell)=0}\,;
\\
&\delta_{--}    &\quad\; & \dom(\delta_{--})=\set{\xi\in\mathrm{W}^1}{ \text{no BV at both $0$ and $\ell$}}=\mathrm{W}^1.
\end{alignat*}

Observe that these four differential operators of first order are auxiliary but necessary for introducing the correct BV from engineering statics for the 
diverse differential operators of 4-th order for the bending beam in the subsequent Section \ref{s3-bending-vibs}.

\begin{lemm}
\label{s3-bending-lemm-1}
The four differential operators $\delta_{.\:.}\in\{\delta_{++},\delta_{+-},\delta_{-+},\delta_{--} \}$ are closed, 
and for their $\mathrm{L}^2$--adjoints it holds
\begin{equation*}
\delta_{++}^*=-\delta_{--}\,,\quad \delta_{--}^*=-\delta_{++}\,,
\qquad\qquad
\delta_{+-}^*=-\delta_{-+}\,,\quad \delta_{-+}^*=-\delta_{+-}\,.
\end{equation*}
\end{lemm}
\begin{proof}
The graph norm of these operators
agrees  with the Sobolev norm $\norm._1$.
So they are closed unbounded operators in $\mathrm{L}^2$ by construction.
Especially, by that construction,
\begin{equation*}
\operatorname{C}_c^\infty((0,\ell))\,,\quad
\left.\operatorname{C}_c^\infty((0,\infty))\right|_{(0,\ell)}\,,\quad
\left.\operatorname{C}_c^\infty((-\infty,\ell))\right|_{(0,\ell)}\,,\quad
\left.\operatorname{C}_c^\infty(\dR)\right|_{(0,\ell)}
\end{equation*}
are cores, which are $\norm._1$--dense in the domains $\dom(\delta_{.\,.})\subseteq\mathrm{W}^1$, respectively.

Let us first consider the pair $\delta_{++}$ and $\delta_{--}$.
According to the construction of the adjoint of an operator in \eqref{s2-def-adjoint} we have
\begin{equation}
\label{eq-s3-bending-lemm-1a}
\dom(\delta_{++}^*)=
\set{\xi \in \mathrm{L}^2}{\exists\,\eta_\xi\in\mathrm{L}^2 \text{\ with\ }
\skal{\eta_\xi}{\vp}= \skal{\xi}{\vp'}\;\:\forall\vp\in \dom(\delta_{++})},
\quad \eta_\xi = \delta_{++}^*\xi\,.
\end{equation}
Suppose first $\xi\in \dom(\delta_{++}^*)$. Applying the testfunctions $\vp\in\operatorname{C}_c^\infty((0,\ell))\subset \dom(\delta_{++})$ to the connection $\skal{\delta_{++}^*\xi}{\vp}=\skal{\xi}{\vp'}=-\skal{\xi'}{\vp}$, we arrive at $\xi\in\mathrm{W}^1=\dom(\delta_{--})$ and $\delta_{++}^*\xi=-\xi'=-\delta_{--}\xi$, especially $\dom(\delta_{++}^*)\subseteq\dom(\delta_{--})=\mathrm{W}^1$.
Conversely, let $\xi\in \mathrm{W}^1$. With partial integration (PI, extension from smooth functions to $\mathrm{W}^1$ via Proposition \ref{s3-prop-W-properties}\ref{s3-prop-W-properties-b}) and $\vp(0)=0=\vp(\ell)$ for $\vp\in\dom(\delta_{++})$ we arrive at 
\begin{equation*}
\skal{\delta_{--}\xi}{\vp}
=
\skal{\xi'}{\vp}
=
\int_0^\ell \overline{\xi'(x)}\vp(x)\,dx
\stackrel{PI}=
\underbrace{\Bigl[ \overline{\xi(x)}\vp(x)\Bigr]_0^\ell}_{=\,0}
-
\int_0^\ell \overline{\xi(x)}\vp'(x)\,dx
=
-\skal{\xi}{\vp'}
\end{equation*}
for all $\vp\in\dom(\delta_{++})$. With help of \eqref{eq-s3-bending-lemm-1a} we conclude that $\xi\in\dom(\delta_{++}^*)$ and $\delta_{++}^*\xi=\eta_\xi=-\xi'=-\delta_{--}\xi$.
Finally, $\delta_{++}^*=-\delta_{--}$.
Adjoining leads to
$\delta_{++}=\delta_{++}^{**}=-\delta_{--}^*$ (note, for every closed operator $B$ it holds $B=B^{**}$).

For the pair $\delta_{+-}$ and $\delta_{-+}$ the situation is different.
By definition of the adjoint it is
\begin{equation}
\label{eq-s3-bending-lemm-1}
\dom(\delta_{-+}^*)
=
\set{\xi \in \mathrm{L}^2}{  \exists\,\eta_{\xi}\in\mathrm{L}^2    \text{\ with\ }
\skal{\eta_{\xi}}{\vp}=\skal{\xi}{\vp'}\;\:\forall \vp\in \dom(\delta_{-+})  },
\quad \eta_\xi = \delta_{-+}^*\xi\,.
\end{equation}
Let first $\xi\in \dom(\delta_{-+}^*)$. Applying the testfunctions $\vp\in\operatorname{C}_c^\infty((0,\ell))\subset \dom(\delta_{-+})$ to the connection $\skal{\delta_{-+}^*\xi}{\vp}=\skal{\xi}{\vp'}=-\skal{\xi'}{\vp}$, we arrive at $\xi\in\mathrm{W}^1$ and $\delta_{-+}^*\xi=-\xi'$. 
In order to prove that $\xi\in\dom(\delta_{+-})$ we integrate partially (PI)
\begin{align*}
\skal{\delta_{-+}^*\xi}{\vp}
&=
-\skal{\xi'}{\vp}
=
-\int_0^\ell \overline{\xi'(x)}\vp(x)\,dx
\stackrel{PI}=
-\Bigl[ \overline{\xi(x)}\vp(x)\Bigr]_0^\ell
+
\int_0^\ell \overline{\xi(x)}\vp'(x)\,dx
\\
&=
\underbrace{-\overline{\xi(\ell)}\vp(\ell)+\overline{\xi(0)}\vp(0)}_{\text{boundary term}} \,+\, \skal{\xi}{\vp'}
\,,
\end{align*}
Thus $\skal{\delta_{-+}^*\xi}{\vp}= \skal{\xi}{\vp'}$ is fulfilled
for all $\vp\in \dom(\delta_{-+})$, if and only if the boundary term vanishs. 
We know $\vp(\ell)=0 $ for all $\vp\in\dom(\delta_{-+})=\{\vp\in\mathrm{W}^1\mid \vp(\ell)=0\}$, but there is no BV at $x=0$ for $\vp\in\dom(\delta_{-+})$.
Thus a vanishing boundary term forces $\xi(0)=0$, implying $\xi\in \dom(\delta_{+-})=\{\xi\in\mathrm{W}^1\mid \xi(0)=0\}$.
So far, $\delta_{-+}^*\xi=-\xi'=-\delta_{+-}\xi$ for $\xi\in\dom(\delta_{-+}^*)\subseteq\dom(\delta_{+-}) $.
Conversely, let $\xi\in\dom(\delta_{+-})$. Then the above PI yields
$\skal{-\delta_{+-}\xi}{\vp}=\skal{-\xi'}{\vp}=\skal{\xi}{\vp'}$ for all $\vp\in\dom(\delta_{-+})$, implying 
$\xi\in\dom(\delta_{-+}^*)$ and $\delta_{-+}^*\xi=\eta_\xi=-\xi'=-\delta_{+-}\xi$ with help of \eqref{eq-s3-bending-lemm-1}.
Therefore, $\delta_{-+}^*=-\delta_{+-}$, and by adjoining $\delta_{-+}=\delta_{-+}^{**}=-\delta_{+-}^*$.
\end{proof}
\begin{rema}[Anti-Selfadjoint Differential Operators of First Order]
\mbox{}\hfill There\\
exist
overcountably many anti-selfadjoint differential operators $\delta_z=-\delta_z^*$ operating in $\mathrm{L}^2$.
For each $z\in\dC$ with $\betr{z}=1$ the operator $\delta_z$ is defined as differentiation of first order with BV different to above,
\begin{equation*}
\delta_z \xi=\xi'\,,\qquad\forall\xi\in\dom(\delta_z):=\{\xi\in \mathrm{W}^{1}\mid \xi(0)=z\xi(\ell)\}\,.
\end{equation*}
So, $\delta_z$ fulfills the boundary condition $\xi(0)=z\xi(\ell)$.
See e.g.\ \cite{ReedSimon}\,Vol.\,I p.\,259, \cite{ReedSimon}\,Vol.\,II p.\,141\,f, or \cite{Weidmann} p.\,240\,f,
additional properties are found in \cite{HoneggerRiekers15}\,Subsection\,17.5.1.

The operators $\delta_z$ are related to the two types $\delta_{++}$ and $\delta_{--}$, in the sense that
\begin{equation*}
\delta_{++}\subset \delta_z\subset\delta_{--}\,,
\quad\text{meaning}\quad
\dom(\delta_{++})\subset \dom(\delta_z)\subset\dom(\delta_{--})\,.
\end{equation*}
$\delta_{++}$ is the smallest, $\delta_{--}$ the largest differential operator in $\mathrm{L}^2$,
whereas all the anti-selfadjoint operators $\delta_z$ lie in between, also $\delta_{+-}$ and $\delta_{-+}$.

Note, when multiplying with $-i$ and with Planck's constant $\hbar$, one arrives  at the
selfadjoint momentum operators $p_z=-i\hbar\delta_z$,
used in one dimensional quantum mechanics on the spatial interval $[0,\ell]$.
\end{rema}
\subsection{Four Different Laplace Operators on the Interval $(0,\ell)$}
\label{ss-3:diff-op-Laplace}
Here for our interval $(0,\ell)$ a Laplace operator $\Delta$ is defined to act by double differentiation as $\Delta\xi=\xi''$ on functions 
$\xi:(0,\ell)\ra\dC$, $ x\mapsto\xi(x)$ of the single variable $x\in(0,\ell)$. Also when multiplying with the minus sign, that is $-\Delta$, the differential operator is denoted a Laplace operator or simply a Laplacian. 
In order to obtain positivity and selfadjointness of $-\Delta$ as an operator acting in $\mathrm{L}^2=\mathrm{L}^2((0,\ell))$ we have to introduce some BV, e.g.\ analogously to Subsection \ref{ss3.1-wavelike}. 

Here, however, we proceed by defining 4 different Laplacians in terms of operator products of suitable differential operators of first order, possessing the desired BV and in addition positivity and selfadjointness.
So let us recall, that the operator product $BC$ of the two operators $B$ and $C$ is defined by
\begin{equation}
\label{s4-eq-operatorproduct}
\dom(BC)=\set{\xi\in\dom(C)}{C\xi\in\dom(B)}\,,
\quad
BC\xi := (BC)\xi=B(C\xi)\,.      
\end{equation}

Since for a closed operator $B$ the operator product $B^*B$ is always positive and selfadjoint,
one immediately obtains the next result with help of Lemma~\ref{s3-bending-lemm-1}.
The indices, $DN$, $DD$, etc., denote homogeneous Dirichlet or Neumann BV
at the left or right boundary point, $x=0$ or $x=\ell$, of the interval $(0,\ell)$, respectively.

\begin{coro}[Four Positive, Selfadjoint Laplacians on $(0,\ell)$]
\label{s3-coro-Laplacians}
\mbox{}\hfill
Consider the\\ following four Laplace operators acting in the Hilbert space $\mathrm{L}^2$ (differentiation in the distributional sense).
\mbox{ }
\begin{enumerate}
\renewcommand{\labelenumi}{(\alph{enumi})}
\renewcommand{\theenumi}{(\alph{enumi})}
\item
The positive, selfadjoint \emph{Dirichlet} Laplacian is given by the operator product
\begin{equation*}
-\Delta_{DD}=\delta_{++}^*\delta_{++}=-\delta_{--}\delta_{++}\,,
\qquad
-\Delta_{DD}\xi=-\xi'',\quad\xi\in\dom(-\Delta_{DD})\subset\mathrm{W}^2.
\end{equation*}
From $\Delta_{DD}=\delta_{--}\delta_{++}$ and the fact that $\delta_{--}$ has no BV,
it follows that $-\Delta_{DD}$ satisfies the BV for $\delta_{++}$, only, namely
the homogeneous Dirichlet BV $\xi(0)=0=\xi(\ell)$.
In the literature this BV commonly is indexed by ``$\infty$'' as $-\Delta_{DD}=-\Delta_\infty$.
\item
The positive, selfadjoint \emph{Neumann} Laplacian is given by the operator product
\begin{equation*}
-\Delta_{NN}=\delta_{--}^*\delta_{--}=-\delta_{++}\delta_{--}\,,
\qquad
-\Delta_{NN}\xi=-\xi'',\quad\xi\in\dom(-\Delta_{NN})\subset\mathrm{W}^2.
\end{equation*}
Since $\delta_{--}$ has no BV, the BV for $-\Delta_{NN}$ arises from the BV of $\delta_{++}$. 
The operator product condition \eqref{s4-eq-operatorproduct} 
implies for the first derivatives $\delta_{--}\xi=\xi'\in\dom(\delta_{++})$,
and hence we arrive at the homogeneous Neumann BV $\xi'(0)=0=\xi'(\ell)$.
In the literature this BV commonly is indexed by a zero ``$0$'', that is $-\Delta_{NN}=-\Delta_0$.
\item
The two positive, selfadjoint \emph{mixed} Laplacians are given by the operator products
\begin{gather*}
-\Delta_{DN}=\delta_{+-}^*\delta_{+-}=-\delta_{-+}\delta_{+-}\,,
\qquad
-\Delta_{DN}\xi=-\xi'',\quad\xi\in\dom(-\Delta_{DN})\subset\mathrm{W}^2,
\\
-\Delta_{ND}=\delta_{-+}^*\delta_{-+}=-\delta_{+-}\delta_{-+}\,,
\qquad
-\Delta_{ND}\xi=-\xi'',\quad\xi\in\dom(-\Delta_{ND})\subset\mathrm{W}^2,
\end{gather*}
with mixed homogeneous Dirichlet and Neumann BV,  $\xi(0)=0=\xi'(\ell)$ and $\xi'(0)=0=\xi(\ell)$, respectively.
\end{enumerate}
Note that by Proposition \ref{s3-prop-W-properties}\ref{s3-prop-W-properties-c} we have $\mathrm{W}^2\subseteq \operatorname{C}^1([0,\ell])$, and thus $\xi$ and $\xi'$ allow for a boundary evaluation, and so the BV are defined as usual.
\end{coro}

The eigenspectra with eigenvalues $a_n$ and \emph{normalized} eigenvectorfunctions $\psi_n$,
\begin{equation*}
-\Delta_{.\:.}\,\psi_n = a_n\psi_n\,,\qquad\forall n\in\dN\,,  
\end{equation*}
of the four Laplacians $-\Delta_{.\:.}$ acting in the Hilbert space $\mathrm{L}^2$
from this Corollary~\ref{s3-coro-Laplacians} are well known to be
\begin{alignat*}{5}
&-\Delta_{DD}=-\delta_{--}\delta_{++}:&
&\;&&
&\quad&
a_n=\bigl({\textstyle\frac{n\pi}{\ell}}\bigr)^2,
&\;\;&
\psi_n(x)=\textstyle\sqrt{{\frac{2}{\ell}}}\,\sin({\textstyle\frac{n\pi}{\ell}}x)\,,\;\;  n\in\dN
\,;
\\
&-\Delta_{NN}=-\delta_{++}\delta_{--}:&
&\;&&
&\quad&
a_n=\bigl({\textstyle\frac{n\pi}{\ell}}\bigr)^2, 
&\;\;&
\psi_n(x)=\textstyle\sqrt{{\frac{2}{\ell}}}\,\cos({\textstyle\frac{(n-1)\pi}{\ell}}x)\,,\;\;  n\geq2
\,,
\\
&&&& &
&\quad& 
a_1=0\,,
&\;\;&
\psi_1(x)=\textstyle{\frac{1}{\sqrt{\smash[b]{\ell}}}}\,,\qquad\qquad\qquad\;\;\;  n=1
\,;
\\
&-\Delta_{DN}=-\delta_{-+}\delta_{+-}:&
&\;&&
&\quad&
a_n=\bigl({\textstyle\frac{(2n-1)\pi}{2\ell}}\bigr)^2\!,
&\;\;&
\psi_n(x)=\textstyle\sqrt{{\frac{2}{\ell}}}\,\sin({\textstyle\frac{(2n-1)\pi}{2\ell}}x)\,,\;\; n\in\dN
\,;
\\
&-\Delta_{ND}=-\delta_{+-}\delta_{-+}:&
&\;&&
&\quad&
a_n=\bigl({\textstyle\frac{(2n-1)\pi}{2\ell}}\bigr)^2\!,
&\;\;&
\psi_n(x)=\textstyle\sqrt{{\frac{2}{\ell}}}\,\cos({\textstyle\frac{(2n-1)\pi}{2\ell}}x)\,,\;\; n\in\dN
\,;
\end{alignat*}
with $x\in[0,\ell]$, of course. (Note, that may be easily derived with help of the theory of Fourier series expansions
by applying certain odd or even symmetry arguments.)
Exclusively the Neumann Laplacian $-\Delta_{NN}$ possesses the eigenvalue zero for which
the constant functions constitute the one-dimensional eigenspace.
In each case the normalized eigenfunctions $\{ \psi_n\in \mathrm{L}^2 \mid n\in\dN\}$
constitute an ONB of the Hilbert space $\mathrm{L}^2=\mathrm{L}^2((0,\ell))$, leading to four different ONBs.

Some of these eigenfunctions are not contained in every domain or range (= image) of the four differential operators
$\delta_{++}$, $\delta_{--}$, $\delta_{-+}$, $\delta_{+-}$ of first order.
For example, for odd $m=2n-1\in\dN$ we obtain
\begin{equation}
\label{s4-eq-domains+ranges}
\begin{aligned}
&\quad\text{it holds\ \ }\sin({\textstyle\frac{m\pi}{2\ell}}x)\in\dom(\delta_{+-})\subset\dom(\delta_{--})\,,\qquad \text{$m$ odd},
\\
&\qquad\text{but\ \ }
\sin({\textstyle\frac{m\pi}{2\ell}}x)\not\in\dom(\delta_{-+})
\text{\ \  and\ \  }
\sin({\textstyle\frac{m\pi}{2\ell}}x)\not\in\dom(\delta_{++})\,,
\\[1ex]
&\delta_{+-}\sin({\textstyle\frac{m\pi}{2\ell}}x)
=
\delta_{--}\sin({\textstyle\frac{m\pi}{2\ell}}x)
=
\textstyle\frac{d}{dx}\sin({\textstyle\frac{m\pi}{2\ell}}x)
=
{\textstyle\frac{m\pi}{2\ell}}\cos({\textstyle\frac{m\pi}{2\ell}}x)\,,
\\[1ex]
&\quad\text{where\ \ }
\cos({\textstyle\frac{m\pi}{2\ell}}x)\in\text{range}(\delta_{+-})\subseteq\dom(\delta_{-+})\subset\dom(\delta_{--})\,,
\\
&\qquad\text{but\ \ }
\cos({\textstyle\frac{m\pi}{2\ell}}x)\not\in\dom(\delta_{+-})
\text{\ \  and\ \  }
\cos({\textstyle\frac{m\pi}{2\ell}}x)\not\in\dom(\delta_{++})\,.
\end{aligned}
\end{equation}
Remark, this fact finally provokes the result that the eigenequation for group \ref{s3-coro-no-Fextension-b} in Corollary \ref{s3-coro-no-Fextension} below is not analytically solvable, see Summary \ref{s4-summ-eigenequation}.

%
%
\section{Bending Vibrations of a Beam in One Spatial Dimension}
\label{s3-bending-vibs}
%
%

\noindent
\begin{minipage}{8cm}
Here we treat free bending vibrations in detail.
Given a slender, isotropic, homogeneous, straight, elastic beam of length $\ell$ with constant cross-sectional area.\\
The $x$--axis is along the neutral fiber of the beam, and the bending deformations $u(t)(x)=u(x,t)$ are vertical (transversal) to the $x$--axis
with positive $u$--direction down as in the figures.\\
It is assumed that the beam is supported only at its ends $x=0$ and $x=\ell$.\\
In the figures beside, only as a first example both ends are supported flexibly, that is,
each support is free to rotate and has no moment resistance.
\end{minipage}
\hfill
\begin{minipage}{7.1cm}
\begin{center}
\includegraphics[width=7.1cm]{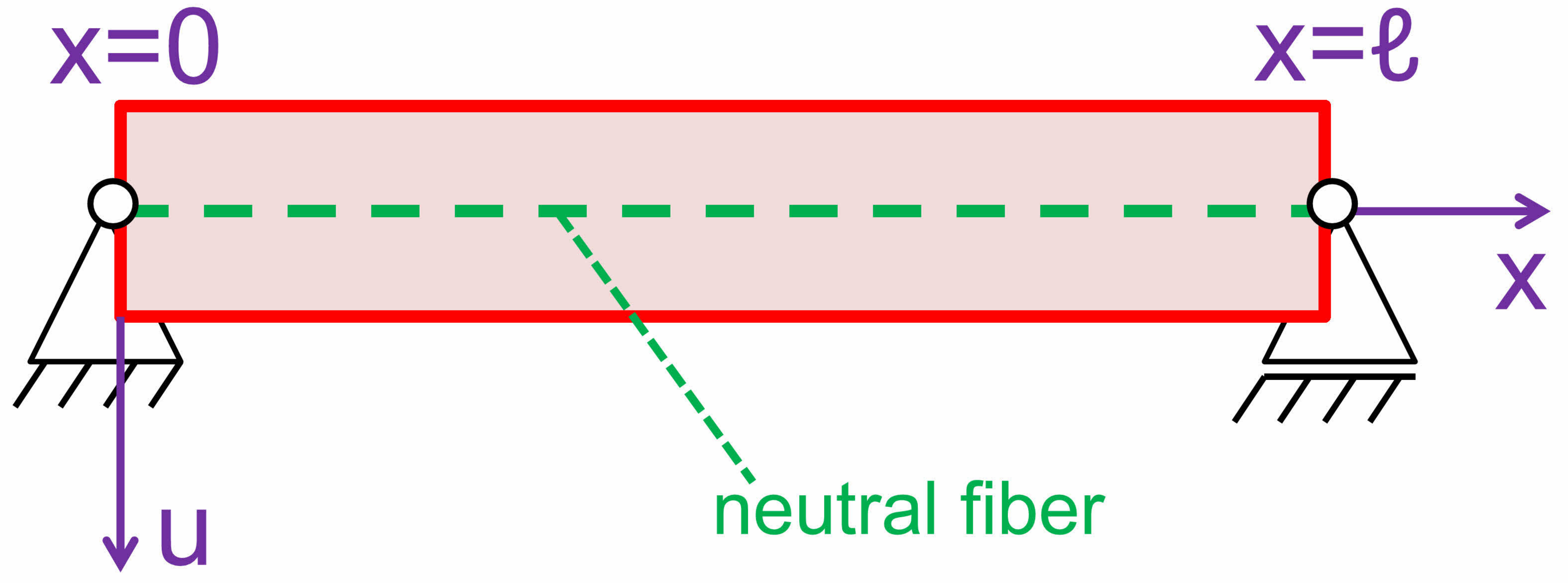}\\
\includegraphics[width=7.0cm]{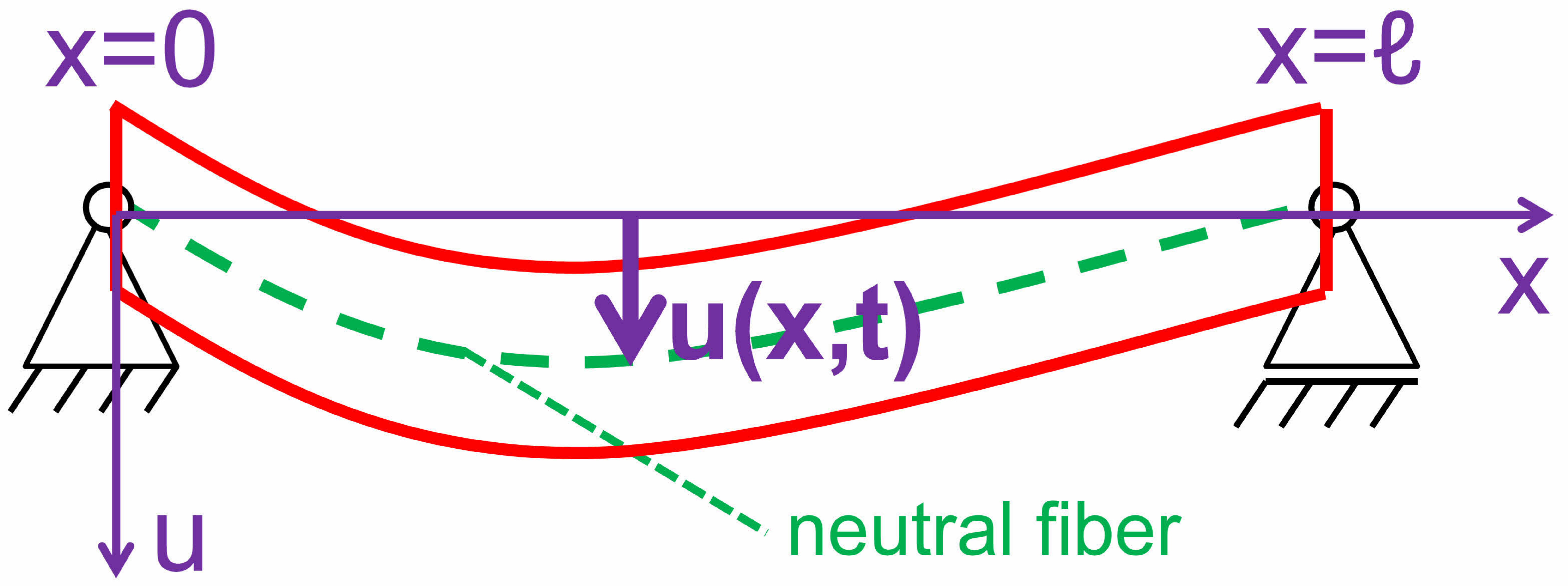}\\
As example, both ends are supported\\ flexibly,
denoted as (a)--(a) support.
\end{center}
\end{minipage}

\vspace{1.1ex}
The beam is along the closed interval $[0,\ell]$ with boundary points $x=0$ and $x=\ell$.
So we deal with spatial functions $\xi:(0,\ell)\ra\dC,\,x\mapsto\xi(x)$ on the open interval $(0,\ell)$, for which we are interested in their behaviour at the two boundary points, where the BV have to be installed.

In the present section we take over our notions introduced in the previous Section~\ref{s3-sobolev+diffOps}. 
Especially we use our complex Hilbert space $\mathrm{L}^2= \mathrm{L}^2((0,\ell))$ of square integrable complex-valued functions on the open interval $(0,\ell)$
for describing the bending vibrations of the beam in terms of the wave-like Euler--Bernoulli differential equation. Nevertheless we are interested in real solutions, only.

\subsection{Differential Operators of 4-th Order for Beam BV}
\label{ss-3:beam+BV}
For beams usually the following three support possibilities are used, well known from statics.
That is, for the transverse, purely spatial bending function $\xi(x)$ we choose at one end $x^\star$ of the beam,
$ x^\star=0$ or $ x^\star=\ell$, the following diverse BV, e.g.\ \cite{MaPoSe} etc.,
\begin{alignat*}{3}
&\text{(a) = flexible support} &\qquad& \xi(x^\star)=0\,, &\qquad& \xi''(x^\star)=0\,;
\\
&\text{(b) =  fixed support} &\qquad& \xi(x^\star)=0\,, &\qquad& \xi'(x^\star)=0\,;
\\
&\text{(c) = free end} &\qquad& \xi''(x^\star)=0\,, &\qquad& \xi'''(x^\star)=0\,.
\end{alignat*}
\begin{note}
\label{s3-note-support=(a)--(c)}
If the left end of the beam, $ x^\star=0$, is supported according to (a) and the right end, $ x^\star=\ell$, is supported by (b), then we briefly call
the beam to be (a)--(b) supported. Analogously, (c)--(b), (b)--(a), $\ldots$ , and so on.
\end{note}

For these different possibilities of support at the ends of the beam we will construct positive, selfadjoint differential operators $A$
of 4-th order, $A\xi=\xi^{(4)}=\xi''''$ (differentiation in the distributional sense), so that the BV are respected.
Moreover, we add three cases, which are not in agreement with the above supports known from statics.

Subsequently we list in the first column the considered support of the beam, then the associated four BV, and finally in the third column 
the corresponding product differential operator $\hat{A}$ of 4-th order respecting exactly these four support BV. 
(See \eqref{s4-eq-operatorproduct} for operator products.)
Finally the positive, selfadjoint operator $A$ turns out to be a unique extension of the product operator $\hat{A}$ 
for each of the beam supports.
\begin{alignat*}{4}
&\text{(a)--(a)}&\text{\ \ \  BV: \ }& \xi(0)=0\,,\;\; \xi(\ell)=0\,,\;\;\xi''(0)=0\,,\;\; \xi''(\ell)=0\,,
&\quad& \hat{A}=\delta_{--}\delta_{++}\delta_{--}\delta_{++}\,;
\\
&\text{(a)--(b)}&\text{\ \ \  BV: \ }& \xi(0)=0\,,\;\; \xi(\ell)=0\,,\;\; \xi'(\ell)=0\,,\;\;\xi''(0)=0\,,
&\quad& \hat{A}=\delta_{--}\delta_{+-}\delta_{-+}\delta_{++}\,;
\\
&\text{(a)--(c)}&\text{\ \ \  BV: \ }& \xi(0)=0\,,\;\; \xi''(0)=0\,,\;\;\xi''(\ell)=0\,,\;\; \xi'''(\ell)=0\,,
&\quad& \hat{A}=\delta_{-+}\delta_{++}\delta_{--}\delta_{+-}\,;
\\
&\text{(b)--(b)}&\text{\ \ \  BV: \ }& \xi(0)=0\,,\;\; \xi(\ell)=0\,,\;\;\xi'(0)=0\,,\;\; \xi'(\ell)=0\,,
&\quad& \hat{A}=\delta_{--}\delta_{--}\delta_{++}\delta_{++}\,;
\\
&\text{(b)--(c)}&\text{\ \ \  BV: \ }& \xi(0)=0\,,\;\; \xi'(0)=0\,,\;\;\xi''(\ell)=0\,,\;\; \xi'''(\ell)=0\,,
&\quad& \hat{A}=\delta_{-+}\delta_{-+}\delta_{+-}\delta_{+-}\,;
\\
&\text{(c)--(c)}&\text{\ \ \  BV: \ }& \xi''(0)=0\,,\;\; \xi''(\ell)=0\,,\;\;\xi'''(0)=0\,,\;\; \xi'''(\ell)=0\,,
&\quad& \hat{A}=\delta_{++}\delta_{++}\delta_{--}\delta_{--}\,;
\\
&\text{add-(i)}&\text{\ \ \  BV: \ }& \xi'(0)=0\,,\;\; \xi'(\ell)=0\,,\;\;\xi'''(0)=0\,,\;\; \xi'''(\ell)=0\,,
&\quad& \hat{A}=\delta_{++}\delta_{--}\delta_{++}\delta_{--}\,;
\\
&\text{add-(ii)}&\text{\ \ \  BV: \ }& \xi(0)=0\,,\;\; \xi'(\ell)=0\,,\;\;\xi''(0)=0\,,\;\; \xi'''(\ell)=0\,,
&\quad& \hat{A}=\delta_{-+}\delta_{+-}\delta_{-+}\delta_{+-}\,;
\\
&\text{add-(iii)}&\text{\ \ \  BV: \ }& \xi(\ell)=0\,,\;\; \xi'(0)=0\,,\;\;\xi''(\ell)=0\,,\;\; \xi'''(0)=0\,,
&\quad& \hat{A}=\delta_{+-}\delta_{-+}\delta_{+-}\delta_{-+}\,.
\end{alignat*}
For the remaining possibilities (b)--(a), (c)--(a), (c)--(b) of statics, simply invert the beam.
Also BV add-(iii) is the inverted beam with BV add-(ii).
Note, we have 
\begin{equation*}
\operatorname{C}_c^\infty((0,\ell))\subset\dom(\hat{A})\subset \mathrm{W}^{4}\,,
\end{equation*}
especially it follows that each $\hat{A}$ is densely defined in
our Hilbert space $\mathrm{L}^2= \mathrm{L}^2((0,\ell))$.
By construction it holds for each of the above product operators $\hat{A}$ that
\begin{equation*}
\hat{A}\xi=\xi'''',\qquad \forall \xi\in
\dom(\hat{A})=\set{\xi \in \mathrm{W}^4}{\xi\text{ fulfills all $4$ BV of $\hat{A}$}}\,.
\end{equation*}
Since $\mathrm{W}^4\subseteq \operatorname{C}^3([0,\ell])$ according to Proposition \ref{s3-prop-W-properties}\ref{s3-prop-W-properties-c}, the boundary evaluations for $\xi$, $\xi'$, $\xi''$, and $\xi'''$ at $x=0$ and $x=\ell$ are well defined.

\begin{note}[Operator Extension]
\label{s5.1-operator-extension}
For an operator $A$ to be an extension of the operator $\hat{A}$,  means
\begin{equation*}
\dom(\hat{A})\subseteq\dom(A)\qquad\text{with}\qquad\hat{A}\xi=A\xi\,,\;\;\forall \xi\in\dom(\hat{A})\,,
\end{equation*}
denoted as $\hat{A}\subseteq A$.
We write $\hat{A}\subset A$ or equivalently $A \supset \hat{A}$, if $A$ is a \emph{genuine} operator extension of $\hat{A}$,
that is with $\dom(\hat{A})\subsetneqq\dom(A)$.
\end{note}
\begin{theo}[Existence of Unique Positive, Selfadjoint Extensions]
\label{s3-theo-pos-sa-extension}
In each of the above support cases, there exists  a unique positive, selfadjoint operator extension
$A\supseteq\hat{A}$  for the product operator $\hat{A}$ such that
\begin{equation}
\label{s4a-condition-Asa}
\dom(A)
\subseteq
\set{\xi \in \mathrm{W}^2}{\xi\text{ fulfills the BV for $\xi$ and $\xi'$ of $\hat{A}$ (not for higher derivatives)}}.
\end{equation}
Moreover, for that unique $A$ it holds,
\begin{align*}
&\: A\xi=\xi'''',\quad \forall \xi\in\dom(A)\cap \mathrm{W}^4.
\\
&\dom(A)\cap\mathrm{W}^4=\set{\xi \in \mathrm{W}^4}{\xi\text{ fulfills all $4$ BV of $\hat{A}$}}=\dom(\hat{A})\,.
\end{align*}
\end{theo}
Its proof is given together with the proof of the subsequent Theorem~\ref{s3-theo-pos-sa-extension-2}
in Section~\ref{s6-proofs}.

$A$ is the so-called \emph{Friedrichs extension} of
the product operator $\hat{A}$, the smallest form extension of $\hat{A}$.
That procedure to obtain a positive, selfadjoint extension operator $A$  of the product operator $\hat{A}$ is based on well known mathematical techniques, e.g., \cite{Wloka82}, \cite{Weidmann} Section 5.5, \cite{Kato84} Chapter six,
\cite{ReedSimon} Vol.\,2 section\,X.3, etc.  

Possibly there may exist further positive, selfadjoint extensions of $\hat{A}$, but then their domains of definition
have to contain elements not from the subspace $\set{\xi \in \mathrm{W}^2}{\xi\text{ fulfills the BV for $\xi$ and $\xi'$ of $\hat{A}$ (not for higher derivatives)}}$, 
meaning, condition \eqref{s4a-condition-Asa} is not fulfilled.

\begin{coro}
\label{s3-coro-no-Fextension}
We distinguish two groups \ref{s3-coro-no-Fextension-a} and \ref{s3-coro-no-Fextension-b} of operators of type
$A$ or $\hat{A}$.
\begin{enumerate}
\renewcommand{\labelenumi}{(\Roman{enumi})}
\renewcommand{\theenumi}{(\Roman{enumi})}
\item
\label{s3-coro-no-Fextension-a}
In each of the four supports (a)--(a), add-(i), add-(ii), and add-(iii), the positive, selfadjoint operator $A$
is \emph{not a genuine} operator extension of $\hat{A}$, since already the operator product in $\hat{A}$
is positive and selfadjoint und so coincides with $A$,
\begin{alignat*}{3}
&\text{(a)--(a)}&
&\qquad&
A=\hat{A}=\delta_{--}\delta_{++}\delta_{--}\delta_{++}=(-\Delta_{DD})^2\,;
\\
&\text{add-(i)}&
&\qquad&
A=\hat{A}=\delta_{++}\delta_{--}\delta_{++}\delta_{--}=(-\Delta_{NN})^2\,;
\\
&\text{add-(ii)}&
&\qquad&
A=\hat{A}=\delta_{-+}\delta_{+-}\delta_{-+}\delta_{+-}=(-\Delta_{DN})^2\,;
\\
&\text{add-(iii)}&
&\qquad&
A=\hat{A}=\delta_{+-}\delta_{-+}\delta_{+-}\delta_{-+}=(-\Delta_{ND})^2\,.
\end{alignat*}
The spectrum of $A$ is the square of the spectrum of the associated positive, selfadjoint Laplace operator from Corollary~\ref{s3-coro-Laplacians}, but of course with the same normalized eigenvectorfunctions $\psi_n$ as stated in Subsection~\ref{ss-3:diff-op-Laplace}.
\item
\label{s3-coro-no-Fextension-b}
In the cases (a)--(b), (a)--(c), (b)--(b), (b)--(c), (c)--(c), the positive, selfadjoint extension $A$ is a \emph{genuine} operator extension of the original product operator $\hat{A}$, that is $A\supset \hat{A}$.
Here each $A$ is defined in terms of a positive closed sesquilinear form (Friedrichs extension).
Therefore, the spectrum of $A$ is not related to the spectra of the four Laplacians from Corollary \ref{s3-coro-Laplacians}.
\end{enumerate}
\end{coro}
\begin{proof}
Part \ref{s3-coro-no-Fextension-a} is an immediate consequence of Subsection~\ref{ss-3:diff-op-Laplace}, since also
the Laplacians are operator products. So $\hat{A}$ is already selfadjoint, and for a selfadjoint operator there do not exist
selfadjoint extensions nor selfadjoint restrictions.\\[-2.6ex]
\indent
We prove part \ref{s3-coro-no-Fextension-b} for the example (a)--(c) with $\hat{A}=\delta_{-+}\overbrace{\delta_{++}\delta_{--}}^{=\,\Delta_{NN}}\delta_{+-}$; all other cases work analogously.
By direct inspection, taking into account the BV as in \eqref{s4-eq-domains+ranges}, it is seen that none of the eigenfunctions
$\sin({\textstyle\frac{m\pi}{\ell}}x)$, $\cos({\textstyle\frac{m\pi}{\ell}}x)$, and 
$\sin({\textstyle\frac{m\pi}{2\ell}}x)$, $\cos({\textstyle\frac{m\pi}{2\ell}}x)$ (with $m$ odd)
of the Laplacians from Subsection~\ref{ss-3:diff-op-Laplace} is contained in
the domain of definition of the product operator $\hat{A}$.
Especially, if $\cos({\textstyle\frac{m\pi}{\ell}}x)\in\dom(\Delta_{NN})$ is an eigenfunction of $\Delta_{NN}$, then
$\cos({\textstyle\frac{m\pi}{\ell}}x)\in\text{range}(\delta_{+-})$, since
$\sin({\textstyle\frac{m\pi}{\ell}}x)\in\dom(\delta_{+-})$.
But $\cos({\textstyle\frac{m\pi}{\ell}}x)\not\in\dom(\delta_{-+})$, and so the last operator $\delta_{-+}$ cannot be applied.
Hence the spectrum of $A$ has nothing to do with the spectrum of any Laplacian.
\end{proof}

Nevertheless, for all cases of group \ref{s3-coro-no-Fextension-b} of the above Corollary \ref{s3-coro-no-Fextension}
one may deduce some general statement concerning
the spectrum of the unique positive, selfadjoint extension operator $A\supseteq\hat{A}$.
Since $\Lam=(0,\ell)$ is interior, one expects a spectral result as stated in Subsection~\ref{s2-spectral-properties}.
\begin{theo}
\label{s3-theo-pos-sa-extension-2}
Let $A\supseteq\hat{A}$ be an above positive, selfadjoint extension for an arbitrarily chosen support in Theorem \ref{s3-theo-pos-sa-extension}
(both groups in Corollary \ref{s3-coro-no-Fextension}).
Then it holds:
\begin{enumerate}
\renewcommand{\labelenumi}{(\alph{enumi})}
\renewcommand{\theenumi}{(\alph{enumi})}
\item
\label{s3-theo-pos-sa-extension-2-(a)}
$A$ has a pure point (= purely discrete) spectrum $\sigma(A)=\sigma_{p}(A)\subset[0,\infty)$.
\item
\label{s3-theo-pos-sa-extension-2-(b)}
Each eigenspace is finite dimensional and the eigenvalues $a_n$, $n\in\dN$,
from \eqref{eq:WEq-2-pps} may be arranged increasingly,
$$
0\leq a_1\leq a_2\leq a_3\leq a_4\leq a_5 \leq .....
$$
according to their finite multiplicity. They converge to infinity,
$\lim\limits_{n\ra\infty} a_n=\infty$.
\item
\label{s3-theo-pos-sa-extension-2-(c)}
Only when $A$ belongs to one of the supports (a)--(c) [also (c)--(a)], (c)--(c), and add-(i), then $A$ possesses the eigenvalue zero.
For (a)--(c) [also (c)--(a)] and add-(i) the eigenspace to the eigenvalue zero is one-dimensional, whereas
for (c)--(c) the eigenspace to the eigenvalue zero is two-dimensional.
\end{enumerate}
\end{theo}
Part \ref{s3-theo-pos-sa-extension-2-(c)} of the Theorem means that rotation of the beam is allowed, and for support
(c)--(c) even transversal translation of the beam is possible.
This corresponds to eigenfunctions of type $\eta(x)=a+bx$ for all $x\in(0,\ell)$ with constants $a,b\in\dR$,
possessing eigenvalue zero since $\eta''=0$.
As mentioned, the proof is found in Section~\ref{s6-proofs}.

\subsection{Euler--Bernoulli Beam Equation in $\mathrm{L}^2((0,\ell))$}
\label{ss-3:PDE-Euler--Bernoulli}
With the previous Subsection \ref{ss-3:beam+BV} the main work is already done. It remains to apply our primary result for the wave-like IVP with purely discrete spectrum for the positive, selfadjoint operator $A$ from Subsection \ref{s1-discrete}, only.

The Euler--Bernoulli (E-B) differential equation IVP for the bending vibrations reads in Hilbert space language
as in Theorem~\ref{theo-1} but with material modification as outlined in Remark \ref{s3-substitute},
\begin{alignat}{2}
\label{eq:WEq-30}
&\text{E-B differential equation}            &\qquad     & \frac{d^2u(t)}{dt^2}=-\varsigma^2Au(t)\,,  \quad t \in\dR\,,
\\   \notag
&\text{IV (at $t=0$)}                        &    & \left.u(t)\right|_{t=0}=u_0  \in\mathrm{L}^2\,,
\\   \notag
&\text{IV (at $t=0$)}                        &    & \left.\frac{du(t)}{dt}\right|_{t=0}  = \dot{u}_0\in\mathrm{L}^2\,,
\end{alignat}
with given IV functions $u_0, \:\dot{u}_0\in\mathrm{L}^2$.
The positive, selfadjoint differential (extension) operator $A\supseteq\hat{A}$ of $4$-th order
has to be chosen according to Theorem \ref{s3-theo-pos-sa-extension} for the considered support,
(a)--(b), (b)--(b), etc., multiplied with the material constant
$$
\varsigma^2:={\frac{E I}{\rho \mathrm{A}_{\text{c-s}}}}\,,
$$
arising from the constant cross-sectional area $\mathrm{A}_{\text{c-s}}$, the mass density per unit length $\rho$,
the elasticity modulus $E$ for the material of the beam, and the second area moment $I$ of the cross-section,
e.g.~\cite{MaPoSe},  etc.    

With the purely discrete spectrum, that is the eigenspectrum of $A$, 
\begin{equation}
\label{ss4a_eigenequation-discrete}
A\psi_n=a_n\psi_n\,,\quad\forall n\in\dN\,,
\end{equation}
with eigenvalues $a_n\geq0$ and 
corresponding normalized eigenvectors $\psi_n\in\mathrm{L}^2$ constituting an ONB,
the $\mathrm{L}^2$--Hilbert space solution trajectory $t\mapsto u(t)$ of the IVP \eqref{eq:WEq-30} is given according to Corollary \ref{coro-1-discrete} 
and Remark \ref{s3-substitute} by the spectral decomposition
\begin{equation}
\label{eq:WEq-sol-2-9}
u(t)
=
\sum_{n=1}^\infty
\Bigl(
\cos(t\varsigma\sqrt{\smash[b]{a_n}})\skal{\psi_n}{u_0}  +
\frac{\sin(t\varsigma\sqrt{\smash[b]{a_n}})}{\varsigma\sqrt{\smash[b]{a_n}}}\skal{\psi_n}{\dot{u}_0}
\Bigr)
\psi_n\:\in\,\mathrm{L}^2\,,
\quad t\in\dR\,.
\end{equation}

%
%
\section{Solving Concretely the Eigenequation}
\label{s4b-eigenequation}
\subsection{The Eigenequation \eqref{ss4a_eigenequation-discrete} as Ordinary Differential Equation}
\label{ss-3:PDE-eigen}
So far we dealed with the abstract results from functional analysis. 
In order to arrive at a concrete solution of the Euler--Bernoulli problem, it is necessary to calculate explicitly the eigenvalues $a_n\geq 0$ and eigenvectors $\psi_n\in\mathrm{L}^2=\mathrm{L}^2((0,\ell))$ for a chosen positive, selfadjoint operator $A$ from the preceding Section \ref{s3-bending-vibs}.

Each positive, selfadjoint differential operator $A$ of group \ref{s3-coro-no-Fextension-b} in Corollary \ref{s3-coro-no-Fextension} is more than only $\frac{d^4}{dx^4}$ on $\mathrm{W}^{4}$ fulfilling the associated BV, but possesses a specific domain of definition containing some elements not from $\mathrm{W}^{4}$.
Nevertheless it seems that here regularity can also be proven, similarly for e.g.\ the Laplace operators in the literature.
And so, without giving a proof, we may assume that the eigenfunctions are sufficiently smooth, at least contained in $\operatorname{C}^4([0,\ell])$.

\begin{obse}[Eigenequation for the Eigenfunctions]
\label{obse-eigenequation}
Under the preceding assumption of sufficiently smooth eigenvectors $\psi_n$ the eigenequation \eqref{ss4a_eigenequation-discrete} is just the linear, homogeneous ordinary differential equation of $4$-th order,
\begin{equation}
\label{eq:WEq-2-pps-9b}
A\psi_n(x)=\psi_n''''(x)=a_n\psi_n(x)\,,\qquad \forall x\in (0,\ell)\,,\qquad a_n\geq 0\,.
\end{equation}
Here for example the (a)--(b) support causes the boundary condition
\begin{equation}
\label{eq:WEq-2-pps-9b-BV}
\text{(a)--(b) \qquad BV:}\quad \psi_n(0)=0\,,\quad \psi_n(\ell)=0\,,\quad \psi'_n(\ell)=0\,,\quad \psi''_n(0)=0\,.
\end{equation}
\end{obse}

The two groups \ref{s3-coro-no-Fextension-a} and \ref{s3-coro-no-Fextension-b} of support cases in Corollary \ref{s3-coro-no-Fextension} 
behave very different in solving the eigenequation. 
\begin{summ}[Solving the Eigenequation]
\label{s4-summ-eigenequation}
The solution of the ordinary differential \eqref{eq:WEq-2-pps-9b} with associated BV 
for supports of group \ref{s3-coro-no-Fextension-a}, namely (a)--(a),  add-(i), add-(ii), and add-(iii), 
is derived in the subsequent Subsection \ref{ss-3:PDE-Euler--Bernoulli-(a)--(a)}.
These are the only analytically solvable cases, they belong
to nonproper operator ``extensions'' but  coincide with the operator products according to
Corollary~\ref{s3-coro-no-Fextension}\ref{s3-coro-no-Fextension-a}, that is $A=\hat{A}$.

For the remaining support cases, group \ref{s3-coro-no-Fextension-b} in Corollary \ref{s3-coro-no-Fextension},
for which the positive, selfadjoint operators $A$ are \emph{genuine} operator extensions of the product operators $\hat{A}$, 
meaning $A\supset\hat{A}$, 
the eigenequation \eqref{eq:WEq-2-pps-9b} with associated BV (like in example \eqref{eq:WEq-2-pps-9b-BV} for (a)--(b) support) is not solvable analytically,  however by numerical methods, only, e.g.\ \cite{MaPoSe}, \cite{Westermann}, and references therein.
The numerics is briefly outlined in Subsection \ref{ss-3:PDE-eigen-numerics} below.
\end{summ}
%
%

%
%
\subsection{The Analytically Solvable Eigenequations}
\label{ss-3:PDE-Euler--Bernoulli-(a)--(a)}
In the four support cases (a)--(a), add-(i), add-(ii), add-(iii) of group \ref{s3-coro-no-Fextension-a} in Corollary~\ref{s3-coro-no-Fextension}
the eigenequation \eqref{eq:WEq-2-pps-9b} is solvable analytically, since
each positive, selfadjoint operator $A=\hat{A}=(-\Delta_{.\:.})^2$ is the square of one of the four Laplacians $-\Delta_{.\:.}$ 
in Subsection \ref{ss-3:diff-op-Laplace}. 
Consequently, by the spectral calculus \eqref{eq:WEq-2-pps-f}
the eigenvalues $a_n$ of $A=(-\Delta_{.\:.})^2$ are the square of the eigenvalues
of the Laplacians, respectively, but with the same normalized eigenvectors $\psi_n$
from Subsection~\ref{ss-3:diff-op-Laplace} (with $x\in[0,\ell]$), leading to
\begin{alignat*}{5}
&\text{(a)--(a)}&
&\;&& A=(-\Delta_{DD})^2:
&\quad&
a_n=\bigl({\textstyle\frac{n\pi}{\ell}}\bigr)^4,
&\;\;&
\psi_n(x)=\textstyle\sqrt{{\frac{2}{\ell}}}\,\sin({\textstyle\frac{n\pi}{\ell}}x)\,,\;\;  n\in\dN
\,;
\\
&\text{add-(i)}&
&\;&& A=(-\Delta_{NN})^2:
&\quad&
a_n=\bigl({\textstyle\frac{n\pi}{\ell}}\bigr)^4, 
&\;\;&
\psi_n(x)=\textstyle\sqrt{{\frac{2}{\ell}}}\,\cos({\textstyle\frac{(n-1)\pi}{\ell}}x)\,,\;\;  n\geq2
\,,
\\
& & & & &
&\quad& 
a_1=0\,,
&\;\;&
\psi_1(x)=\textstyle{\frac{1}{\sqrt{\smash[b]{\ell}}}}\,,\qquad\qquad\qquad\;\;\,  n=1
\,;
\\
&\text{add-(ii)}&
&\;&& A=(-\Delta_{DN})^2:
&\quad&
a_n=\bigl({\textstyle\frac{(2n-1)\pi}{2\ell}}\bigr)^4\!,
&\;\;&
\psi_n(x)=\textstyle\sqrt{{\frac{2}{\ell}}}\,\sin({\textstyle\frac{(2n-1)\pi}{2\ell}}x)\,,\;\; n\in\dN
\,;
\\
&\text{add-(iii)}&
&\;&& A=(-\Delta_{ND})^2:
&\quad&
a_n=\bigl({\textstyle\frac{(2n-1)\pi}{2\ell}}\bigr)^4\!,
&\;\;&
\psi_n(x)=\textstyle\sqrt{{\frac{2}{\ell}}}\,\cos({\textstyle\frac{(2n-1)\pi}{2\ell}}x)\,,\;\; n\in\dN
\,.
\end{alignat*}

For each of these four supports the unique solution $u(t)$ of the Euler--Bernoulli beam IVP is given by equation \eqref{eq:WEq-sol-2-9}.
For almost all $x\in [0,\ell]$ and $t\in\dR$ one arrives at
\begin{equation}
\label{eq:WEq-sol-2-9-b}
u(x,t)=u(t)(x)
=
\sum_{n=1}^\infty
\!
\Bigl(
\cos(t\varsigma\sqrt{\smash[b]{a_n}})\skal{\psi_n}{u_0}  +
\frac{\sin(t\varsigma\sqrt{\smash[b]{a_n}})}{\varsigma\sqrt{\smash[b]{a_n}}}\skal{\psi_n}{\dot{u}_0}
\Bigr)
\psi_n(x)\,.
\end{equation}

Note, it arises from the investigation of the spectra of the four Laplace operators in Subsection \ref{ss-3:diff-op-Laplace} that only for BV case add-(i) with $A=(-\Delta_{NN})^2$ one obtains the eigenvalue zero, namely $a_1=0$ for $n=1$. In accordance with equation~\eqref{s2-eq-sin(ty)/y} 
for that eigenvalue $a_1=0$ we have 
\begin{equation*}
\cos(t\varsigma\sqrt{\smash[b]{a_1}})\stackrel{a_1=0}=\cos(0)=1\,,
\qquad
\frac{\sin(t\varsigma\sqrt{\smash[b]{a_1}})}{\varsigma\sqrt{\smash[b]{a_1}}}\stackrel{a_1=0}=t\,.
\end{equation*}

\subsection{Reporting on the Numerics of the Eigenequation}
\label{ss-3:PDE-eigen-numerics}
Let us briefly outline the numerics of the eigenequation from Observation \ref{obse-eigenequation}, but for the strictly positive eigenvalues $a_n>0$, only,
cf.\ \cite{MaPoSe}, \cite{Westermann}, and references therein.
The eigenvalue zero is treated in Theorem \ref{s3-theo-pos-sa-extension-2}\ref{s3-theo-pos-sa-extension-2-(c)}.
The general solution of the linear ordinary differential equation of $4$-th order occurring in the eigenequation \eqref{eq:WEq-2-pps-9b}, 
\begin{equation*}
\psi_n''''(x)=a_n\psi_n(x)\,,\qquad a_n>0\,, 
\end{equation*}
is well known to be given by
\begin{equation}
\label{s4-eq-eigenfunction-Mpsin}
\psi_n(x)
=
\alpha_{1,n}\cosh(\kappa_n x)+\alpha_{2,n}\sinh(\kappa_n x)+\alpha_{3,n}\cos(\kappa_n x)+\alpha_{4,n}\sin(\kappa_n x)
\end{equation}
with arbitrary constants $\alpha_{1,n},\ldots,\alpha_{4,n}\in\dR$,  where 
\begin{equation*}
\kappa_n := \sqrt[4]{a_n}\:>\,0\,.
\end{equation*}

As example let us take the (a)--(b) support, not possessing the eigenvalue zero. The other support cases work analogously. The (a)--(b) BV in \eqref{eq:WEq-2-pps-9b-BV} yield
\begin{alignat*}{5}
0\;=\;\psi_n(0)&\;=\;\bigl[\alpha_{1,n}&&     &\:+\:&\alpha_{3,n}&&&&\bigr]\,,
\\
0\;=\;\psi_n''(0)&\;=\;\bigl[\alpha_{1,n}&& &\:-\:&\alpha_{3,n}&&&&\bigr]\kappa_n^2\,,
\\
0\;=\;\psi_n(\ell)&\;=\;\bigl[\alpha_{1,n}\cosh(\kappa_n \ell)&\:+\:&\alpha_{2,n}\sinh(\kappa_n \ell)&\:+\:&\alpha_{3,n}\cos(\kappa_n \ell)&\:+\:&\alpha_{4,n}\sin(\kappa_n \ell)&&\bigr]\,,
\\
0\;=\;\psi_n'(\ell)&\;=\;\bigl[\alpha_{1,n}\sinh(\kappa_n \ell)&\:+\:&\alpha_{2,n}\cosh(\kappa_n \ell)&\:-\:&\alpha_{3,n}\sin(\kappa_n \ell)&\:+\:&\alpha_{4,n}\cos(\kappa_n \ell)&&\bigr]\kappa_n\,.
\end{alignat*}
In other words, the BV rewrite as the system of linear equations
\begin{equation}
\label{s4-eq-eigenvalue-Matrix}
\begin{pmatrix}
0\\0\\0\\0
\end{pmatrix}
=
\underbrace{
\begin{pmatrix}
1&0&1&0
\\
1&0&-1&0
\\
\cosh(\kappa_n \ell)&\sinh(\kappa_n \ell)&\cos(\kappa_n \ell)&\sin(\kappa_n \ell)
\\
\sinh(\kappa_n \ell)&\cosh(\kappa_n \ell)&-\sin(\kappa_n \ell)&\cos(\kappa_n \ell)
\end{pmatrix}
}_{\mbox{$=M$}}
\begin{pmatrix}
\alpha_{1,n}\\\alpha_{2,n}\\\alpha_{3,n}\\\alpha_{4,n}
\end{pmatrix}.
\end{equation}
In order to obtain a nontrivial function $\psi_n(x)$ in \eqref{s4-eq-eigenfunction-Mpsin}, the vector 
$\left(\begin{smallmatrix}
\alpha_{1,n}\\\alpha_{2,n}\\\alpha_{3,n}\\\alpha_{4,n}
\end{smallmatrix}\right)\neq 0$
should be nontrivial.
That is only possible when the determinant of the matrix $M$
vanishes. It is immediately calculated that $\det(M)=0$ is equivalent to
\begin{equation}
\label{s4-eq-eigenvalue-kappan=an4}
\tanh(\kappa_n \ell)=\tan(\kappa_n \ell)\,.
\end{equation}
The latter commonly is called the \emph{eigenvalue equation} for the (a)--(b) support, too. 
From the solutions $\kappa_n>0$ of \eqref{s4-eq-eigenvalue-kappan=an4}
one derives the eigenvalues $a_n$ of the eigenequation \eqref{eq:WEq-2-pps-9b} by
\begin{equation*}
a_n=\kappa_n^4\,>\,0\,,\qquad\text{eigenvalue to $A$ in \eqref{eq:WEq-2-pps-9b} for (a)--(b) support}.
\end{equation*}
In accordance with Theorem \ref{s3-theo-pos-sa-extension-2}
the values $\kappa_n= \sqrt[4]{a_n}$ may be arranged increasingly, 
\begin{equation*}
0< \kappa_1\leq \kappa_2\leq \kappa_3\leq \kappa_4\leq \kappa_5 \leq .....
\,,\qquad\quad
\lim_{n\ra\infty} \kappa_n=\infty\,.
\end{equation*}
The described procedure ensures that each eigenspace of $A$ is one dimensional.

(a)--(b) support belongs to group \ref{s3-coro-no-Fextension-b} in Corollary \ref{s3-coro-no-Fextension}, and hence the eigenvalue equation \eqref{s4-eq-eigenvalue-kappan=an4} is solvable numerically, only. A so found numerical $\kappa_n$ then has to be inserted into the linear system \eqref{s4-eq-eigenvalue-Matrix} in order to obtain suitable vectors $\left(\begin{smallmatrix}
\alpha_{1,n}\\\alpha_{2,n}\\\alpha_{3,n}\\\alpha_{4,n}
\end{smallmatrix}\right)\neq 0$,
from which one finally arrives at a normalized eigenfunction $\psi_n(x)$ via equation \eqref{s4-eq-eigenfunction-Mpsin}, being indeed infinitely smooth.

Here at our example for (a)--(b) support the first and second line in \eqref{s4-eq-eigenvalue-Matrix} imply $\alpha_{1,n}=0=\alpha_{3,n}$. So the system of linear equation \eqref{s4-eq-eigenvalue-Matrix} reduces to 
\begin{equation*}
\begin{pmatrix}
0\\0
\end{pmatrix}
=
\begin{pmatrix}
\sinh(\kappa_n \ell)&\sin(\kappa_n \ell)
\\
\cosh(\kappa_n \ell)&\cos(\kappa_n \ell)
\end{pmatrix}
\begin{pmatrix}
\alpha_{2,n}\\ \alpha_{4,n}
\end{pmatrix}.
\end{equation*}
(Note, also the vanishing determinant here is equivalent to the eigenvalue equation \eqref{s4-eq-eigenvalue-kappan=an4}.) Putting $\alpha_n:=\alpha_{4,n}$ one gets
$\alpha_{2,n}=-\frac{\cos(\kappa_n \ell)}{\cosh(\kappa_n \ell)}\alpha_n=-\frac{\sin(\kappa_n \ell)}{\sinh(\kappa_n \ell)}\alpha_n$. Thus by \eqref{s4-eq-eigenfunction-Mpsin},
\begin{equation}
\label{s4-eq-eigenfunction-Mpsin-(a)--(b)}
\psi_n(x)
=
\alpha_{n}\Bigl(-\frac{\cos(\kappa_n \ell)}{\cosh(\kappa_n \ell)}\sinh(\kappa_n x)+\sin(\kappa_n x)\Bigr)\,,
\qquad x\in[0,\ell]\,,
\end{equation}
where $\alpha_n\in\dR$ has to be chosen such that 
\begin{equation*}
\norm{\psi_n}^2=\int_0^\ell |\psi_n(x)|^2dx=1\,.
\end{equation*}

In order to obtain a numerical solution of the Euler--Bernoulli IVP, one has to use the above solution formula \eqref{eq:WEq-sol-2-9}. However, numerics forces the infinite sum there to be approximately reduced to a finite sum ranging only over some finite number $N$ of additive terms, depending on the desired degree of numeric approximation,
\begin{equation}
\label{eq:WEq-sol-2-9-numeric=N}
u(x,t)=u(t)(x)
=
\sum_{n=1}^N
\Bigl(
\cos(t\varsigma\kappa_n^2)\skal{\psi_n}{u_0}  +
\frac{\sin(t\varsigma\kappa_n^2)}{\varsigma\kappa_n^2}\skal{\psi_n}{\dot{u}_0}
\Bigr)
\psi_n(x)
\end{equation}
for $x\in[0,\ell]$ and $t\in\dR$. Of course, both occurring scalar products, 
\begin{equation}
\label{ss4b-scalar-prods}
\skal{\psi_n}{u_0}
=
\int_0^\ell \overline{\psi_n(x)} u_0(x)\,dx
\,,\qquad
\skal{\psi_n}{\dot{u}_0}
=
\int_0^\ell \overline{\psi_n(x)} \dot{u}_0(x)\,dx
\,,
\end{equation}
also may be calculated numerically.


%
%
\section{Wave Swinging of a String versus Bending Vibrations of a Beam}
\label{s4-Weq-EB-wave}
In this section we compare the solution of a wave equation with the solution for the beam bending equation derived in the prior section. 
Directly comparable are homogeneous Dirichlet BV for the wave IVP and the support case (a)--(a) for the Euler--Bernoulli bending equation, only.

\subsection{String Wave Swinging with Homogeneous Dirichlet BV}
The IVP for the {\bf wave equation} in one spatial dimension with homogeneous Dirichlet BV is given in function language as
\begin{alignat*}{3}
&\text{wave PDE}                     &\qquad & \partial_t^2 u = c^2 \partial_x^2u = - c^2 (-\partial_x^2)u\,,  \quad && x\in(0,\ell)\,,\quad t \in\dR\,, \\
&\text{IV (at $t=0$)}                        &    & u(x,0)  = u_0(x)\,, \quad && x\in(0,\ell)\,,    \\
&\text{IV (at $t=0$)}                        &    & \partial_t u(x,0)  = \dot{u}_0(x)\,, \quad && x\in[0,\ell]\,,  \\
&\text{hom.~Dirichlet BV}                 &     & u(0,t)  = 0\,, \quad  u(\ell,t) = 0 \,,\quad && t \in\dR\,,
\end{alignat*}
with wave speed $c>0$, length $\ell>0$, and  $u_0(x),\:\dot{u}_0(x)$ are two given IV functions.
The  solution can be interpreted as a swinging string (of a violin or guitar) with length $\ell$,
which is fixed at both ends.
Translated into Hilbert space language, we arrive at
\begin{alignat}{2}
\label{eq:WEq-50}
&\text{wave differential equation}           &\qquad     & \frac{d^2u(t)}{dt^2}=-c^2(\overbrace{-\Delta_{DD}}^{=\,A})u(t)\,,
\quad t \in\dR\,,
\\   \notag
&\text{IV (at $t=0$)}                        &    & \left.u(t)\right|_{t=0}=u_0  \in\mathrm{L}^2\,,
\\   \notag
&\text{IV (at $t=0$)}                        &    & \left.\frac{du(t)}{dt}\right|_{t=0}  = \dot{u}_0\in\mathrm{L}^2\,,
\end{alignat}
with given IV functions $u_0, \:\dot{u}_0\in\mathrm{L}^2$. Note, in Hilbert space terminology the BV are covered by the positive, selfadjoint Dirichlet Laplace operator $A=-\Delta_{DD}$ from Corollary~\ref{s3-coro-Laplacians}.

The normalized eigenfunctions $\psi_n$ of $A=-\Delta_{DD}$ with corresponding eigenvalues $a_n>0$ are stated in Subsection \ref{ss-3:diff-op-Laplace}, namely 
\begin{equation}
\label{s5:eigenspectrum=-DeltaDD}
\psi_n(x)=\textstyle\sqrt{{\frac{2}{\ell}}}\,\sin({\textstyle\frac{n\pi}{\ell}}x)\,,\;\;x\in [0,\ell]\,,
\quad\text{with eigenvalue}\quad
a_n=\bigl({\textstyle\frac{n\pi}{\ell}}\bigr)^2,\quad \forall n\in\dN\,.
\end{equation}

According to Corollary \ref{coro-1-discrete} and Remark \ref{s3-substitute}
the unique solution of the wave IVP is given in Hilbert space language as
\begin{equation}
\label{s5:eq-sol-wave}
u(t)
=
\sum_{n=1}^\infty
\Bigl(
\cos(tc\sqrt{\smash[b]{a_n}})\skal{\psi_n}{u_0}  +
\frac{\sin(tc\sqrt{\smash[b]{a_n}})}{c\sqrt{\smash[b]{a_n}}}\skal{\psi_n}{\dot{u}_0}
\Bigr)
\psi_n\,,
\quad\forall t\in\dR\,.
\end{equation}

The inner products of the eigenvectors $\psi_n$ with the IV functions $u_0,\:\dot{u}_0 \in\mathrm{L}^2$,
\begin{equation}
\label{s3-WEq-eigen-Laplace-EF}
\begin{split}
\skal{\psi_n}{u_0}
&=
\sqrt{\frac{2}{\ell}}\int_0^\ell u_0(x)\sin({\textstyle\frac{n\pi}{\ell}}x)\,dx
=:
\sqrt{\frac{\ell}{2}}\:S_n
\,,
\\
\skal{\psi_n}{\dot{u}_0}
&=
\sqrt{\frac{2}{\ell}}\int_0^\ell \dot{u}_0(x)\sin({\textstyle\frac{n\pi}{\ell}}x)\,dx
=:
\sqrt{\frac{\ell}{2}}\:\dot{S}_n
\,,
\end{split}
\end{equation}
appearing in the solution \eqref{s5:eq-sol-wave}, constitute just the sine Fourier coefficients $S_n$ and $\dot{S}_n$
of the (odd extensions to $(-\ell,\ell)$ of the) IV functions $u_0,\: \dot{u}_0\in\mathrm{L}^2$, respectively.
Using these coefficients the wave solution \eqref{s5:eq-sol-wave} rewrites as an ordinary (smooth) function by
\begin{equation}
\label{s5:eq-sol-wave-funct}
u(t)(x)=u(x,t)
=
\sum_{n=1}^\infty
\Bigl[
S_n \cos(\om_n t)+\frac{\dot{S}_n}{\om_n} \sin(\om_n t)
\Bigr]
\sin({\textstyle\frac{n\pi}{\ell}}x)
\end{equation}
for almost all $x\in [0,\ell]$ and all $t\in\dR$, where
\begin{equation*}
\om_n=c\sqrt{\smash[b]{a_n}}=\frac{n \pi}{\ell}\:c\,,
\qquad\forall n\in\dN\,.
\end{equation*}
\eqref{s5:eq-sol-wave-funct} is just
the sine Fourier series expansion (with respect to the spatial variable $x$).

\subsection{Beam Bending Vibrations with (a)--(a) Support}
We take the (a)--(a) operator $A=(-\Delta_{DD})^2$ for the {\bf Euler--Bernoulli differential equation}
from \eqref{eq:WEq-30}.
The normalized eigenfunctions $\psi_n$ of $A=(-\Delta_{DD})^2$ with corresponding eigenvalues $a_n>0$ are stated in Subsection \ref{ss-3:PDE-Euler--Bernoulli-(a)--(a)}, namely 
$$
\psi_n(x)=\textstyle\sqrt{{\frac{2}{\ell}}}\,\sin({\textstyle\frac{n\pi}{\ell}}x)\,,\;\;x\in [0,\ell]\,,
\quad\text{with eigenvalue}\quad
a_n=\bigl({\textstyle\frac{n\pi}{\ell}}\bigr)^4,\quad \forall n\in\dN\,.
$$
Recall, by spectral calculus \eqref{eq:WEq-2-pps-f} the eigenfunctions $\psi_n$ are the same as for the Laplacian $-\Delta_{DD}$ 
from \eqref{s5:eigenspectrum=-DeltaDD} but with square for the eigenvalues of $-\Delta_{DD}$.

With the sine Fourier coefficients $S_n$ and $\dot{S}_n$
as defined in equation \eqref{s3-WEq-eigen-Laplace-EF} above,  
the unique solution formula \eqref{eq:WEq-sol-2-9} or \eqref{eq:WEq-sol-2-9-b} of the Euler--Bernoulli IVP 
turns out to agree with the sine Fourier series expansion (with respect to the spatial variable $x$)
\begin{equation}
\label{s5:eq-sol-E-B-funct}
u(t)(x)=u(x,t)
=
\sum_{n=1}^\infty
\Bigl[
S_n \cos(\om_n t)+\frac{\dot{S}_n}{\om_n} \sin(\om_n t)
\Bigr]
\sin({\textstyle\frac{n\pi}{\ell}}x)
\end{equation}
for almost all $x\in [0,\ell]$ and all $t\in\dR$, where here
\begin{equation*}
\om_n=\varsigma\sqrt{\smash[b]{a_n}}=\Bigl({\frac{n\pi}{\ell}}\Bigr)^2\varsigma\,,
\qquad\forall n\in\dN\,.
\end{equation*}

For completeness let us state the Euler--Bernoulli IVP in function language. With ordinary solution function $u(t)(x)=u(x,t)$ and
BV support (a)--(a), it writes as
\begin{alignat*}{3}
&\text{Euler--Bernoulli PDE}               &     & \partial_t^2u = - \varsigma^2 \partial_x^4 u\,,  \quad && x\in(0,\ell)\,,\quad t \in\dR\,, \\
&\text{IV (at $t=0$)}                        &    & u(x,0)  = u_0(x)\,, \quad && x\in(0,\ell)\,,    \\
&\text{IV (at $t=0$)}                        &    & \partial_t u (x,0)  =\dot{u}_0(x)\,, \quad && x\in[0,\ell]\,,  \\
&\text{hom.~Dirichlet BV}\qquad       &     & u(0,t)  = 0\,, \quad  u(\ell,t) = 0 \,,\quad && t \in\dR\,,\\
&\text{hom.~second order BV}\qquad &   &  \partial_x^2u(0,t)  = 0\,, \quad \partial_x^2u(\ell,t) = 0 \,,\quad && t \in\dR\,.
\end{alignat*}

\subsection{Comparison: Wave IVP and Euler--Bernoulli PDE}
Both solutions, wave and Euler--Bernoulli (E-B), look very similar, cf.\ the formulas \eqref{s5:eq-sol-wave-funct} and \eqref{s5:eq-sol-E-B-funct},
provided identical IV functions $u_0,\: \dot{u}_0\in\mathrm{L}^2$ and thus
identical Fourier coefficients $S_n$ and $\dot{S}_n$ by \eqref{s3-WEq-eigen-Laplace-EF}, because of identical normalized eigenfunctions $\psi_n$.

The difference becomes visible from the exponent $j=1$ or $j=2$ in
$A=(-\Delta_{DD})^j$, leading to different eigenvalues and different circular frequencies, 
\begin{alignat*}{4}
&\text{wave:}
&\;\;& A=-\Delta_{DD}\,,
&\;\;&\text{eigenvalue\ \ }a_n=\bigl({\textstyle\frac{n\pi}{\ell}}\bigr)^2,
&\;\;&\text{circular\,frequency\ \ }\om_n=\textstyle\frac{n \pi}{\ell}\:c\,;
\\
&\text{E-B:}
&\;\;& A=(-\Delta_{DD})^2,
&\;\;&\text{eigenvalue\ \ }a_n=\bigl({\textstyle\frac{n\pi}{\ell}}\bigr)^4,
&\;\;&\text{circular\,frequency\ \ }\om_n=\bigl({\textstyle\frac{n\pi}{\ell}}\bigr)^2\varsigma\,.
\end{alignat*}

For each $n\in\dN$ the circular frequency $\om_n$ in the bending case of Euler--Bernoulli we may decompose similarly to the string wave swinging as
\begin{alignat}{3}
\notag
&\text{wave:}
&\qquad& 
\om_n=\frac{n \pi}{\ell}\:c
&\qquad\text{with velocity}\qquad&
c>0\,;
\\
\label{ss6.3-EB-wavespeed}
&\text{E-B:}
&\qquad& 
\om_n
=
\Bigl({\frac{n\pi}{\ell}}\Bigr)^2\varsigma
= 
\frac{n\pi}{\ell}\: c_n
&\qquad\text{with velocity}\qquad&
c_n:=\frac{n\pi}{\ell}\:\varsigma\,.
\end{alignat}
Now the solutions of both IVP are rewritten as
\begin{alignat}{2}
\label{s6-compare-sol-wave}
&\text{wave:}
&\quad& 
u(t)(x)=u(x,t)
=
\sum_{n=1}^\infty
\Bigl[
S_n \cos({\textstyle\frac{n\pi}{\ell}}\,c\, t)+\frac{\dot{S}_n}{{\textstyle\frac{n\pi}{\ell}}\,c} \sin({\textstyle\frac{n\pi}{\ell}}\,c\, t)
\Bigr]
\sin({\textstyle\frac{n\pi}{\ell}}x)\,;
\\
\label{s6-compare-sol-E-B}
&\text{E-B:}
&\quad& 
u(t)(x)=u(x,t)
=
\sum_{n=1}^\infty
\Bigl[
S_n \cos({\textstyle\frac{n\pi}{\ell}}\,c_n t)+\frac{\dot{S}_n}{{\textstyle\frac{n\pi}{\ell}}\,c_n} \sin({\textstyle\frac{n\pi}{\ell}}\,c_n t)
\Bigr]
\sin({\textstyle\frac{n\pi}{\ell}}x)\,.
\end{alignat}

Let us decompose both solutions into left and right propagating waves (running in negative and positive $x$-direction, respectively)
for every $n$-th spatial mode $\sin({\textstyle\frac{n\pi}{\ell}}x)$,
$n\in\dN$, respectively.
This may be immediatley done by use of the trigonometric addition theorems
(for $\alpha=\frac{n\pi}{\ell}x$ and $\beta=\om_n t=\frac{n\pi}{\ell}c_{(n)} t$),
\begin{equation*}
\sin(\alpha)\cos(\beta)
=
\textstyle\frac12\bigl[\sin(\alpha+\beta)+\sin(\alpha-\beta)\bigr]
,\quad
\sin(\alpha)\sin(\beta)
=
\textstyle\frac12\bigl[\cos(\alpha-\beta)-\cos(\alpha+\beta)\bigr].
\end{equation*}
This finally leads to the mode decompositions
\begin{alignat*}{2}
&\text{wave:} &\qquad u(x,t)=\:& 
\frac12
\sum_{n=1}^\infty
\Bigl[
\underbrace{
A_n \sin\bigl({\textstyle\frac{n\pi}{\ell}}(x+c t)\bigr)  -  B_n \cos\bigl({\textstyle\frac{n\pi}{\ell}}(x+c t)\bigr)
}_{\mbox{wave running to the left with velocity $c$}}
\\
&&&
\qquad\qquad
+
\underbrace{
A_n \sin\bigl({\textstyle\frac{n\pi}{\ell}}(x-c t)\bigr)+B_n \cos\bigl({\textstyle\frac{n\pi}{\ell}}(x-c t)\bigr)
}_{\mbox{wave running to the right with velocity $c$}}
\Bigr]
\,;
\\
&\text{E-B:} &\qquad u(x,t)=\:& 
\frac12
\sum_{n=1}^\infty
\Bigl[
\underbrace{
A_n \sin\bigl({\textstyle\frac{n\pi}{\ell}}(x+c_n t)\bigr)  -  B_n \cos\bigl({\textstyle\frac{n\pi}{\ell}}(x+c_n t)\bigr)
}_{\mbox{wave running to the left with velocity $c_n$}}
\\
&&&
\qquad\qquad
+
\underbrace{
A_n \sin\bigl({\textstyle\frac{n\pi}{\ell}}(x-c_n t)\bigr)+B_n \cos\bigl({\textstyle\frac{n\pi}{\ell}}(x-c_n t)\bigr)
}_{\mbox{wave running to the right with velocity $c_n$}}
\Bigr]\,.
\end{alignat*}

For the string wave swinging the velocity of the wave just is the ``$c>0$'' occurring in the wave differential equation \eqref{eq:WEq-50}.
In formula \eqref{s6-compare-sol-wave} this \emph{identical} wave speed $c$ appears in all the partial waves for all the $n$-th modes. 

Concerning Euler--Bernoulli (E-B) the $n$-th mode represents just a wave with wave speed $c_n$. 
This means, the Euler--Bernoulli solution \eqref{s6-compare-sol-E-B} is an infinite sum over all the $n$-th mode waves, not with the same but with \emph{different} wave velocities $c_n=\frac{n\pi}{\ell}\,\varsigma$, depending linearly on $n$ as outlined in equation \eqref{ss6.3-EB-wavespeed}.


%
%
\section{Proof of Theorems \ref{s3-theo-pos-sa-extension} and \ref{s3-theo-pos-sa-extension-2} in Subsection \ref{ss-3:beam+BV}}
\label{s6-proofs}
\begin{small}
\noindent
\textsc{Part\,(a).}
By \cite{Leis86} Theorem 2.6(2) there exists a constant $c>0$ so that
\begin{equation*}
\norm{\xi'}\leq c \bigl(\norm{\xi}+\norm{\xi''}\bigr)\,,\quad\forall\xi\in \mathrm{W}^2.
\end{equation*}
Consequently, we arrive at the following estimate for the Sobolev norm $\norm._2$
(recall, $\norm.$ is the $\mathrm{L}^2$--norm)
\begin{align*}
\norm{\xi}^2_2
&=
\norm{\xi}^2+\norm{\xi'}^2+\norm{\xi''}^2
\\
&\leq
\norm{\xi}^2+c^2 \bigl(\norm{\xi}+\norm{\xi''}\bigr)^2+\norm{\xi''}^2
\\
&=
(1+c^2)\norm{\xi}^2+(1+c^2)\norm{\xi''}^2+2c^2\norm{\xi}\norm{\xi''}
\\
&\stackrel{\star}\leq
(1+c^2)\norm{\xi}^2+(1+c^2)\norm{\xi''}^2+c^2\bigl(\norm{\xi}^2+\norm{\xi''}^2\bigr)
\\
&=
(1+2c^2)\bigl(\underbrace{\norm{\xi}^2+\norm{\xi''}^2}_{\mbox{$=:\norm{\xi}_s^2$}}\bigr)
\\
&\leq
(1+2c^2)\bigl(\norm{\xi}^2+\norm{\xi'}^2+\norm{\xi''}^2\bigr)
\\
&=
(1+2c^2)\norm{\xi}_2^2.
\end{align*}
At the inequality with star $\stackrel{\star}\leq$ we used $0\leq (a-b)^2=a^2+b^2-2ab$, therefore  $2ab\leq a^2+b^2$.
That means, the norm $\norm._s$ and the second Sobolev norm $\norm._2$ are equivalent on $\mathrm{W}^2$.

For the demonstration how to proceed, let us take for example the support case (a)--(b);
the other cases work analogously.
For (a)--(b) it is $\hat{A}=\delta_{--}\delta_{+-}\delta_{-+}\delta_{++}$.
Taking adjoints according to Lemma~\ref{s3-bending-lemm-1} one gets
\begin{equation*}
\skal{\xi}{\hat{A}\eta}=\skal{\delta_{-+}\delta_{++}\xi}{\delta_{-+}\delta_{++}\eta}\,,
\quad \forall \xi\in\dom(\delta_{-+}\delta_{++})\,,\quad\forall \eta\in\dom(\hat{A})\subseteq\mathrm{W}^4.
\end{equation*}
We define the positive sesquilinear form $s$
\begin{align*}
&s(\xi,\eta):=\skal{\delta_{-+}\delta_{++}\xi}{\delta_{-+}\delta_{++}\eta}\,,
\\
&\forall \xi,\eta\in\dom(s):=\dom(\delta_{-+}\delta_{++})=
\set{\eta\in\mathrm{W}^2}{\eta(0)=0\,,\;\;\eta(\ell)=0\,,\;\;\eta'(\ell)=0}
\end{align*}
(since $\mathrm{W}^2\subseteq \operatorname{C}^1([0,\ell])$ the boundary terms are well defined).
Because of the equivalence of the norm $\norm._s$ and the second Sobolev norm $\norm._2$,
it follows that the form $s$ is closed, since $\set{\xi\in\mathrm{W}^2}{\xi(0)=0\,,\;\;\xi(\ell)=0\,,\;\;\xi'(\ell)=0}$
is a $\norm._2$--closed subspace of the second Sobolev space $\mathrm{W}^2$.
Especially it follows that the product operator $\delta_{-+}\delta_{++}$ is closed.

We now cite  \cite{Kato84} Subsection VI \S\,2,\,1 
with a result, which is valid for every positive closed form in any real or complex Hilbert space $\cH$.
\begin{theo}[First Representation Theorem of \cite{Kato84}]
For the positive closed form $s$ there exists a positive, selfadjoint operator $A$ acting in $\cH$
(here $\cH=\mathrm{L}^2$), such that:
\begin{enumerate}
\renewcommand{\labelenumi}{(\roman{enumi})}
\renewcommand{\theenumi}{(\roman{enumi})}
\item
\label{FRT(i)}
$\dom(A)\subseteq\dom(s)$, and
\begin{equation}
\label{s4-eq-first-representation-theorem}
s(\xi,\eta)=\skal{\xi}{A\eta}\,,
\quad \forall \xi\in\dom(s)\,,\quad\forall \eta\in\dom(A)\,.
\end{equation}
\item
\label{FRT(ii)}
$\dom(A)$ is a form core for $s$.
\item
\label{FRT(iii)}
If for $\eta\in\dom(s)$ and $\vp\in \cH$ it holds
$s(\xi,\eta)=\skal{\xi}{\vp}$ for all $\xi$ from a form core of $s$, then
$\eta\in\dom(A)$ and $A\eta=\vp$.
\end{enumerate}
Moreover, uniqueness of $A$ is given by (i).
\end{theo}

Suppose $\xi\in\mathrm{W}^2$ and $\eta\in \mathrm{W}^4$. Then two times integrating partially leads to
(extension from smooth functions by Proposition \ref{s3-prop-W-properties}\ref{s3-prop-W-properties-b})
\begin{equation}
\label{s6-PI+BV}
\skal{\xi''}{\eta''}
=
\Bigl[ \overline{\xi'}\eta''\Bigr]_0^\ell
-\Bigl[ \overline{\xi}\eta'''\Bigr]_0^\ell
+
\skal{\xi}{\eta''''}
\,.
\end{equation}
Recall $\mathrm{W}^m\subseteq \operatorname{C}^k([0,\ell])$ for $m>k$ from Proposition \ref{s3-prop-W-properties}\ref{s3-prop-W-properties-c}, and hence the boundary terms are well defined.
Inserting $\xi\in\dom(s)$ and $\eta\in\mathrm{W}^4\cap\dom(s)$ in \eqref{s6-PI+BV},  it follows that
\begin{equation}
\label{s6-PI+BV-xtra}
\skal{\xi''}{\eta''}
=
- \,\overline{\xi'(0)}\eta''(0)
+
\skal{\xi}{\eta''''}
\,.
\end{equation}
Consequently we arrive at the equivalence
\begin{equation*}
s(\xi,\eta)=\skal{\xi''}{\eta''}=\skal{\xi}{\eta''''}\;\;\forall \xi\in\dom(s)
\qquad\quad \Leftrightarrow \qquad\quad
\eta''(0)=0\,.
\end{equation*}
From the condition $\eta\in\mathrm{W}^4\cap\dom(s)$ together with $\eta''(0)=0$
we conclude from (iii) of the first representation Theorem that $\eta\in\dom(A)$ and $A\eta=\eta''''$.
In other words,
\begin{equation*}
\dom(\hat{A})
=
\set{\eta\in\mathrm{W}^4}{\eta(0)=0\,,\;\;\eta(\ell)=0\,,\;\;\eta'(\ell)=0\,,\;\;\eta''(0)=0}
\subseteq
\dom(A)\,,
\end{equation*}
and consequently, $A$ is an extension of $\hat{A}=\delta_{--}\delta_{+-}\delta_{-+}\delta_{++}$.

If conversely, $\eta\in\dom(A)\cap\mathrm{W}^4$, then doubled partial integration \eqref{s6-PI+BV-xtra} for all $\xi\in\dom(s)$,
\begin{equation*}
s(\xi,\eta)=\skal{\xi''}{\eta''}
=
- \,\overline{\xi'(0)}\eta''(0)
+
\skal{\xi}{\eta''''}
=
- \,\overline{\xi'(0)}\eta''(0)
+
\skal{\xi}{A\eta}
\,,
\end{equation*}
compared with \eqref{s4-eq-first-representation-theorem} ensures $\eta''(0)=0$. Thus
\begin{equation*}
\dom(\hat{A})
=
\set{\eta\in\mathrm{W}^4}{\eta(0)=0\,,\;\;\eta(\ell)=0\,,\;\;\eta'(\ell)=0\,,\;\;\eta''(0)=0}
=
\dom(A)\cap\mathrm{W}^4 .
\end{equation*}

With Example 2.13 of \cite{Kato84} Subsection\,VI\,\S\,2,\,4  
one concludes that
\begin{equation}
\label{s4-eq-form=s-9}
A=(\delta_{-+}\delta_{++})^*\delta_{-+}\delta_{++}\,.
\end{equation}

So far we have proven Theorem \ref{s3-theo-pos-sa-extension} up to the stated uniqueness.

\vspace{2ex}
\noindent
\textsc{Part\,(b).}
Let us turn to derive the stated uniqueness.

We start with support (a)--(a) to be given. Then the  equation analogous to \eqref{s4-eq-form=s-9} is
\begin{equation*}
A=\hat{A}=
\underbrace{\delta_{--}\delta_{++}}_{=\,\Delta_{DD}}
\underbrace{\delta_{--}\delta_{++}}_{=\,\Delta_{DD}}
=(-\Delta_{DD})^2
\end{equation*}
with the positive, selfadjoint Laplacian $-\Delta_{DD}$, in accordance
with Corollary \ref{s3-coro-no-Fextension}\ref{s3-coro-no-Fextension-a}.
Since for a selfadjoint operator there do not exist
proper selfadjoint restrictions nor proper selfadjoint extensions, uniqueness of $A$
is already achieved. The same holds for the other support cases in group \ref{s3-coro-no-Fextension-a},
and so we may exclude subsequently the group
\ref{s3-coro-no-Fextension-a}, namely the support cases (a)--(a), add-(i), add-(ii), and add-(iii). 
But in the following argumentation an exclusion of group \ref{s3-coro-no-Fextension-a} is not necessary.

Let us select now $\hat{A}$ to correspond to any support case.
First note that the above form $s$ is a closed extension of the positive form
\begin{equation*}
\hat{s}(\xi,\eta):=\skal{\xi}{\hat{A}\eta}\,,
\quad
\xi,\eta\in\dom(\hat{s})
:=\dom(\hat{A})
=\set{\xi \in \mathrm{W}^4}{\xi\text{ fulfills all $4$  BV  of $\hat{A}$}}\,.
\end{equation*}
For the detailed proof that $s$ is indeed the smallest closed extension of $\hat{s}$ see \textsc{part\,(d)}.
The heuristics behind that is explained here:
First remember, the norm $\norm._s$ is equivalent to the second Sobolev norm $\norm._2$ by \textsc{part\,(a)},
so the $\norm._s$--closure of  $\dom(\hat{s})$ coincides with its closure with respect to $\norm._2$
within $\mathrm{W}^2$.
Now take into account the fact that $\xi\in\mathrm{W}^2$ does not possess a boundary evaluation for $\xi''$ and $\xi'''$,
only for $\xi$ and $\xi'$, in accordance with Proposition \ref{s3-prop-W-properties}\ref{s3-prop-W-properties-c}.
So, when performing the $\norm._2$--closure of $\dom(\hat{s})=\dom(\hat{A})$ within $\mathrm{W}^2$,
the BV for $\xi''$ and $\xi'''$ of $\xi\in \dom(\hat{A})$ are no longer respected, and
the $\norm._2$--closure of $\dom(\hat{A})$ should agree with
\begin{equation*}
\dom(s)=
\set{\xi \in \mathrm{W}^2}{\xi\text{ fulfills the BV for $\xi$ and $\xi'$ of $\hat{A}$ (but not for higher derivatives)}}.
\end{equation*}
Consequently, $s$ is the smallest closed extension, i.e.\ the closure, of the positive form $\hat{s}$,
and $A$ is the Friedrichs extension of $\hat{A}$, \cite{Kato84} Subsection\,VI\,\S\,2,\,3. 

For the proof of uniqueness of $A$, assume that $\breve{A}\supseteq\hat{A}$ is any positive, selfadjoint extension of $\hat{A}$. Then the corresponding
positive form $\breve{s}(\xi,\eta)=\skal{\xi}{\breve{A}\eta}$, $\xi,\eta\in\dom(\breve{s}):=\dom(\breve{A})$ is closable,
its closure be denoted by the same symbol $\breve{s}$.
The operator $\breve{A}$ is the operator associated to $\breve{s}$ by the first representation Theorem,
Corollary 2.2 of
\cite{Kato84} Subsection\,VI\,\S\,2,\,1,  
that is,
\begin{equation*}
\breve{s}(\xi,\eta)=\skal{\xi}{\breve{A}\eta}\,,
\quad \forall \xi\in\dom(\breve{s})\,,\quad\forall \eta\in\dom(\breve{A})\subseteq\dom(\breve{s})\,.
\end{equation*}
Since $s$ is the smallest closed form (its closure) extending $\hat{s}$,
one concludes $\hat{s}\subseteq s\subseteq\breve{s}$.
If $\dom(\breve{A})\subseteq\dom(s)\subseteq\dom(\breve{s})$, then by restriction to $\dom(s)$,
\begin{equation*}
s(\xi,\eta)=\breve{s}(\xi,\eta)=\skal{\xi}{\breve{A}\eta}\,,
\quad \forall \xi\in\dom(s)\subseteq\dom(\breve{s})\,,\quad\forall
\eta\in\dom(\breve{A})\subseteq\dom(s)\subseteq\dom(\breve{s})\,.
\end{equation*}
According to the uniqueness stated in (i) in the first representation Theorem,
it follows $\breve{A}=A$ and $\breve{s}=s$.
So, $A$ is the unique positive, selfadjoint extension of $\hat{A}$, which fulfills the stated property
$\dom(A)\subseteq\dom(s)=
\set{\xi \in \mathrm{W}^2}{\xi\text{ fulfills the BV for $\xi$ and $\xi'$ of $\hat{A}$ (not higher)}}$.

Up to the proof of the above more intuitive argument given in \textsc{part\,(d)},  Theorem~\ref{s3-theo-pos-sa-extension} now has been proved.

\vspace{2ex}
\noindent
\textsc{Part\,(c).}
We prove here the spectral results of Theorem~\ref{s3-theo-pos-sa-extension-2}.
The identical embeddings $\mathrm{W}^{2}\hookrightarrow\mathrm{W}^1 \hookrightarrow \mathrm{L}^2$
are compact by Proposition \ref{s3-prop-W-properties}\ref{s3-prop-W-properties-d}.
The proven equivalence of norms now ensures that
$(\dom(s),\norm._s)\hookrightarrow\mathrm{L}^2$ is compact,
and \ref{s3-theo-pos-sa-extension-2-(a)} and \ref{s3-theo-pos-sa-extension-2-(b)}
of Theorem~\ref{s3-theo-pos-sa-extension-2} follow e.g.\ from
\cite{HoneggerRiekers15} Proposition 43.5-11, a result outlined also in many further textbooks.
\begin{rema}
Compact embedding arguments for a discrete spectrum (like here) also take place for
the four Laplacians in Subsection~\ref{ss-3:diff-op-Laplace}, but since we are in the interval $(0,\ell)$ 
with two completely smooth boundary points, 
their spectra may be calculated directly with associated smooth eigenvectorfunctions, as we have done there.
\end{rema}
We turn to Theorem~\ref{s3-theo-pos-sa-extension-2}\ref{s3-theo-pos-sa-extension-2-(c)}.
$\eta$ contained in the kernel of $A$ means $\eta\in\dom(A)$ with $A\eta=0$ (kernel = eigenspace to eigenvalue zero),
which leads to $0=\skal{\eta}{A\eta}=s(\eta,\eta)=\norm{\eta''}^2$.
Thus $\eta''=0$. This is a vanishing second distributional derivative, so we may conclude $\eta$ to be of type $\eta(x)=a+bx$ with some constants $a,b\in\dC$.
(This conclusion would not be true, if $\eta''(x)=0$ is valid for almost all $x\in(0,\ell)$, only, without the knowledge
of being a second \emph{distributional} derivative, which is ensured by $\eta$ being an element of $\mathrm{W}^2$.)
Up to (a)--(c), (c)--(c), and add-(i), the other support possibilities imply $a=b=0$ and consequently $\eta=0$.
Inserting the supports (a)--(c), (c)--(c), or add-(i)
into $\eta(x)=a+bx$ finally proves (c)
(if one of the constants $a$ and $b$ is freely to choose, then the kernel of $A$ is one-dimensional, if both, then
two-dimensional).
For (c)--(a) invert the beam.

\vspace{2ex}
\noindent
\textsc{Part\,(d).}
We use the Poincar\'{e} estimate as proven in \cite{Leis86} Section 2.3:   
There exists a constant $k>0$ such that
\begin{equation}
\label{s6-PI-1}
\norm{\xi}
\leq
k
\bigl(\norm{\xi'}+ \betr{\skal{1}{\xi}}\bigr)
\,,
\quad\forall \xi\in\mathrm{W}^1\,.
\end{equation}
Here $\skal{1}{\xi}=\int_0^\ell \xi(x)\,dx$ is the inner product of $\xi$ with the constant unit function
$1(x)=1$ for all $x\in(0,\ell)$. Applying \eqref{s6-PI-1} to $\xi'$ yields
\begin{equation}
\label{s6-PI-2}
\norm{\xi'}
\leq
k
\bigl(\norm{\xi''}+ \betr{\skal{1}{\xi'}}\bigr)
\,,
\quad\forall \xi\in\mathrm{W}^2\,.
\end{equation}
Recall, $\xi \in\mathrm{W}^2\subseteq \operatorname{C}^1([0,\ell])$ is
continuously differentiable, and $\xi''$ is defined in the distributional sense.
Then for all $\xi\in\mathrm{W}^2$,
\begin{align*}
\norm{\xi}^2_s
&\;\,=\;\,
\norm{\xi}^2+\norm{\xi''}^2
\\
&\stackrel{\eqref{s6-PI-1}}\leq
k^2
\bigl[\norm{\xi'}+ \betr{\skal{1}{\xi}}\bigr]^2+\norm{\xi''}^2
\\
&\stackrel{\eqref{s6-PI-2}}\leq
k^2
\bigl[
k
\bigl(\norm{\xi''}+ \betr{\skal{1}{\xi'}}\bigr)+ \betr{\skal{1}{\xi}}\bigr]^2+\norm{\xi''}^2
\\
&\;\,\leq\;\ldots\leq\;
d\bigl(\norm{\xi}^2+\norm{\xi'}^2+\norm{\xi''}^2\bigr)
=
d \norm{\xi}^2_2
\end{align*}
with some constant $d>0$.
For the latter inequality  one has to use the estimate $\betr{\skal{1}{\eta}}\leq\norm{1}\norm{\eta}$
and inequalities like $2ab\leq a^2+b^2$ as in \textsc{part\,(a)}.
It follows that the norm
\begin{equation*}
\norm{\xi}^2_t
:=
\norm{\xi''}^2+ \betr{\skal{1}{\xi'}}^2+ \betr{\skal{1}{\xi}}^2
\end{equation*}
is a third norm on $\mathrm{W}^2$ being equivalent to the norm $\norm._s$ and the second Sobolev norm $\norm._2$.
The associated inner product reads
\begin{equation*}
\skal{\xi}{\eta}_t
=
\skal{\xi''}{\eta''}+ \skal{\xi'}{1}\skal{1}{\eta'}+  \skal{\xi}{1}\skal{1}{\eta}\,,
\quad \forall \xi,\eta\in\mathrm{W}^2.
\end{equation*}

Suppose that the closure $\overline{\hat{s}}$ of the form
\begin{equation*}
\hat{s}(\xi,\eta)=\skal{\xi}{\hat{A}\eta}\,,
\quad \xi,\eta\in\dom(\hat{s})=\dom(\hat{A})
\end{equation*}
from \textsc{part\,(b)} does not agree with the closed, positive form $s$
from \textsc{part\,(a)}.
That is, we have the proper form inclusion
$\overline{\hat{s}}\subset s$, or equivalently,
$\dom(\overline{\hat{s}})$ is a proper closed subspace of $\dom(s)$ with respect to
the equivalent norms $\norm._t\bowtie\norm._s\bowtie\norm._2$ on $\mathrm{W}^2$.
Then there exists a 
\begin{equation*}
\vartheta\in\dom(s)=\set{\vartheta\in \mathrm{W}^2}{\vartheta\text{ fulfills the BV for $\vartheta$ and $\vartheta'$ of $\hat{A}$
                                                                                                (but not for higher derivatives)}},
\end{equation*}
which is orthogonal to $\dom(\overline{\hat{s}})$
with respect to the inner product $\skal.._t$, meaning
\begin{equation}
\label{s6-orthogonal-rel}
\begin{aligned}
&0\;=\;\skal{\xi}{\vartheta}_t
\;=
\underbrace{\skal{\xi''}{\vartheta''}}_{\mbox{$=\skal{\xi''''}{\vartheta}$}}
+\;\: \skal{\xi'}{1}\skal{1}{\vartheta'}\,+\,  \skal{\xi}{1}\skal{1}{\vartheta}\,,
\\
&\text{for all $\xi\in\dom(\overline{\hat{s}})$, or equivalently,}
\\
&\text{for all $\xi$ from the form core
$\;\dom(\hat{s})=\set{\xi \in \mathrm{W}^4}{\xi\text{ fulfills all 4 BV of $\hat{A}$}}$}\,.
\end{aligned}
\end{equation}
Note that the identity
$\skal{\xi''}{\vartheta''}=\skal{\xi''''}{\vartheta}$
is valid only, when $\xi$ is taken from the form core $\dom(\hat{s})\subset \mathrm{W}^4$
of its closure $\overline{\hat{s}}$. This is shown with a double partial integration
analogously to \eqref{s6-PI+BV} taking into account the BV of $\xi\in\dom(\hat{s})$ and of $\vartheta\in\dom(s)$.
\begin{lemm}
\label{s6-lemm-Hilbert}
Let $\phi\in\mathrm{L}^2$. Then the following assertions are valid:
\begin{enumerate}
\renewcommand{\labelenumi}{(\alph{enumi})}
\renewcommand{\theenumi}{(\alph{enumi})}
\item
\label{s6-lemm-Hilbert-a}
If $0=\skal{\xi'}{\phi}$ for all $\xi\in\operatorname{C}_c^\infty((0,\ell))$,
then $\phi(x)=a$ for $x\in (0,\ell)$ with a constant $a\in\dC$.
\item
\label{s6-lemm-Hilbert-b}
Suppose there exists a constant $\beta\in\dC$ such that
\begin{equation*}
0=\skal{\xi''}{\phi}-2\beta\skal{\xi}{1}\,,
\quad\forall\xi\in\operatorname{C}_c^\infty((0,\ell))\,.
\end{equation*}
Then there exist constants $a,b\in\dC$ with
$\phi(x)=a+bx+\beta x^2$ for $x\in(0,\ell)$.
\end{enumerate}
\end{lemm}
In the more general context of distribution theory, \ref{s6-lemm-Hilbert-a} is well known as Hilbert's lemma.
%
\begin{prooff}
Fix a $\vp_0\in \operatorname{C}_c^\infty((0,\ell))$
with $-1=\skal{1}{\vp_0}=\int_0^\ell \vp_0(y)\,dy$.
For each $\xi\in\operatorname{C}_c^\infty((0,\ell))$ define
\begin{equation*}
\psi(x):=
\int_{0}^x \bigl(\xi(y)+\skal{1}{\xi}\vp_0(y)\bigr)dy\,,\quad \forall x\in (0,\ell)\,.
\end{equation*}
Then $\psi\in\operatorname{C}_c^\infty((0,\ell))$ with compact support contained in
$\supp(\vp_0)\cup\,\supp(\xi)$. It holds
\begin{equation*}
\psi'(x)=
\xi(x)+\skal{1}{\xi}\vp_0(x)\,,\qquad
\psi''(x)=
\xi'(x)+\skal{1}{\xi}\vp_0'(x)\,,
\qquad \forall x\in (0,\ell)\,.
\end{equation*}
\ref{s6-lemm-Hilbert-a}
Inserting  $\psi\in\operatorname{C}_c^\infty((0,\ell))$ yields
\begin{equation*}
0=\skal{\psi'}{\phi}
=
\skal{\xi}{\phi}+\skal{\xi}{1}\underbrace{\skal{\vp_0}{\phi}}_{\mbox{$=:\,-a$}}
=
\skal{\xi}{\phi}-\skal{\xi}{a}
=
\skal{\xi}{\phi-a}\,.
\end{equation*}
Since $\xi$ may be chosen arbitrarily and $\operatorname{C}_c^\infty((0,\ell))$ is dense in $\mathrm{L}^2$,
it follows $\phi-a=0$.

\vspace{2ex}
\noindent
\ref{s6-lemm-Hilbert-b}
Double partial integration (PI) leads to
\begin{equation*}
0=
\skal{\xi''}{\phi}-2\beta\skal{\xi}{1}
\stackrel{PI}=
\skal{\xi''}{\phi}-\beta\skal{\xi''}{x^2}
=
\skal{\xi''}{\underbrace{\phi-\beta x^2}_{\mbox{$=:\,\tilde{\phi}$}}}\,.
\end{equation*}
Inserting  $\psi\in\operatorname{C}_c^\infty((0,\ell))$ yields
\begin{equation*}
0=\skal{\psi''}{\tilde{\phi}}
=
\skal{\xi'}{\tilde{\phi}}+\skal{\xi}{1}\underbrace{\skal{\vp_0'}{\tilde{\phi}}}_{\mbox{$=:b$}}
=
\skal{\xi'}{\tilde{\phi}}+\skal{\xi}{b}
\stackrel{PI}=
\skal{\xi'}{\tilde{\phi}}-\skal{\xi'}{bx}
=
\skal{\xi'}{\tilde{\phi}-bx}\,.
\end{equation*}
So by \ref{s6-lemm-Hilbert-a}, $\tilde{\phi}-bx=a$, thus $\phi=a +bx+\beta x^2$.
\end{prooff}
Restricting the orthogonality relation \eqref{s6-orthogonal-rel} to the testfunctions $\xi\in\operatorname{C}_c^\infty((0,\ell))$, we get
\begin{equation*}
0=
\skal{\xi''}{\vartheta''}
+\skal{1}{\vartheta'} \underbrace{\skal{\xi'}{1}}_{=0}  +   \skal{1}{\vartheta}\skal{\xi}{1}
=
\skal{\xi''}{\vartheta''}+   \underbrace{\skal{1}{\vartheta}}_{-2\beta}\skal{\xi}{1}\,,
\end{equation*}
since $\skal{1}{\xi'}=\int_0^\vartheta\xi'(x)\,dx=\xi(\ell)-\xi(0)=0$ because of the compact support of $\xi$.
Then part \ref{s6-lemm-Hilbert-b} of the previous lemma implies
\begin{equation}
\label{s6-eq-eta-sd=}
\vartheta''=a+bx-\frac{\skal{1}{\vartheta}}{2} x^2\,,
\qquad x\in (0,\ell),
\end{equation}
an identity being valid in the distributional or $\mathrm{L}^2$--sense, since $\vartheta\in\mathrm{W}^2$.

In terms of testfunctions with their compact supports,  it is not possible to specify $\vartheta$ in further details,
one has to take the BV of a beam support into account.
Let us reduce the orthogonality relation \eqref{s6-orthogonal-rel} to boundary terms.
This has to be done for every case of $\hat{A}$ or $\hat{s}$ separately.
Moreover, for convenience we set from now on $\ell:=1$  without restriction of generality.
As example we choose the support case (b)--(c),
\begin{equation*}
\text{(b)--(c) with BV for all $\xi\in\dom(\hat{s})$: \ }\xi(0)=0\,,\;\; \xi'(0)=0\,,\;\;\xi''(1)=0\,,\;\; \xi'''(1)=0\,,
\end{equation*}
and for the above $\vartheta$ the condition $\vartheta\in\dom(s)$ yields the BV $\vartheta(0)=0$ and $\vartheta'(0)=0$.
Inserting \eqref{s6-eq-eta-sd=} leads with the BV $\xi(0)=0$ and $\xi'(0)=0$ for $\xi$ and doubled partial integration (PI) to
\begin{align*}
\skal{\xi''}{\vartheta''}
&=
\skal{\xi''}{a+bx-\textstyle\frac{\skal{1}{\vartheta}}{2} x^2}
\\
&
\stackrel{PI}=
\Bigl[\overline{\xi'(x)}\bigl(a+bx-\textstyle\frac{\skal{1}{\vartheta}}{2} x^2\bigr)\Bigr]_0^1
-
\Bigl[\overline{\xi(x)}\bigl(b-\skal{1}{\vartheta}x\bigr)\Bigr]_0^1
-
\skal{\xi}{1}\skal{1}{\vartheta}
\\
&=
\overline{\xi'(1)}\bigl(a+b-\textstyle\frac{\skal{1}{\vartheta}}{2}\bigr)
-
\overline{\xi(1)}\bigl(b-\skal{1}{\vartheta}\bigr)
-
\skal{\xi}{1}\skal{1}{\vartheta}\,.
\end{align*}
Noting $\skal{\xi'}{1}=\int_0^1\overline{\xi'(x)}\, dx=\overline{\xi(1)}$ and $\skal{1}{\vartheta'}=\int_0^1\vartheta'(x)\, dx=\vartheta(1)$
by the BV, now the orthogonality relation \eqref{s6-orthogonal-rel} reads as
\begin{equation*}
0=
\overline{\xi'(1)}\bigl(\underbrace{a+b-\textstyle\frac{\skal{1}{\vartheta}}{2}}_{=\,0}\bigr)
+
\overline{\xi(1)}\bigl(\underbrace{\vartheta(1)-b+\skal{1}{\vartheta}}_{=\,0}\bigr)\,,
\qquad
\forall\xi\in\dom(\hat{s})\,.
\end{equation*}
The expressions in the round brackets vanish because of the following reason:
By the boundary extension Theorem, e.g.~\cite{FischerKaul98} \S\,14, 6.6,  
to all given BV $\kappa^{(m)}(0)$ and $\kappa^{(n)}(1)$  there exists a function
$\kappa\in \operatorname{C}^\infty([0,1])\subset \mathrm{W}^{4}$ satisfying the specified BV.
That means, when varying $\xi$ in $\dom(\hat{s})$, then $\xi'(1)$ and $\xi(1)$ take arbitrary values
independently of each other, and so these expressions have to vanish,
\begin{align}
\label{s6-LGS-1}
&
2a+2b-\skal{1}{\vartheta}=0\,,
\\
&
\label{s6-LGS-2}
-b+\skal{1}{\vartheta}+\vartheta(1)=0\,.
\end{align}
On the other hand, \eqref{s6-eq-eta-sd=} implies with the BV $\vartheta(0)=0$ and $\vartheta'(0)=0$ that
\begin{equation}
\label{s6-LGS-4a}
\vartheta(x)=\frac{a}{2}x^2+\frac{b}{6}x^3-\frac{\skal{1}{\vartheta}}{24} x^4\,,\qquad x\in [0,\ell=1]\,.
\end{equation}
As a first consequence we get by inserting the right boundary point $x=\ell=1$ into \eqref{s6-LGS-4a} (and multiplying by $24$) that
\begin{equation}
\label{s6-LGS-3}
-12a-4b+\skal{1}{\vartheta}+24\vartheta(1)=0\,.
\end{equation}
And when integrating \eqref{s6-LGS-4a} over $[0,1]$ and factor out $\int_0^1\vartheta(x)\,dx=\skal{1}{\vartheta}$ one arrives at the second consequence,
\begin{equation}
\label{s6-LGS-4}
-20a-5b+121\skal{1}{\vartheta}=0\,.
\end{equation}
The formulas \eqref{s6-LGS-1}, \eqref{s6-LGS-2}, \eqref{s6-LGS-3}, and \eqref{s6-LGS-4} perform the system
of linear equations
\begin{equation*}
\begin{pmatrix}
2&2&-1&0
\\
0&-1&1&1
\\
-12&-4&1&24
\\
-20&-5&121&0
\end{pmatrix}
\begin{pmatrix}
a \\ b \\ \skal{1}{\vartheta} \\ \vartheta(1)
\end{pmatrix}
=
\begin{pmatrix}
0\\0\\0\\0
\end{pmatrix},
\end{equation*}
which is uniquely, thus trivially solvable because of a nonzero determinant, meaning
\begin{equation*}
0=a=b=\skal{1}{\vartheta}=\vartheta(1)\qquad\stackrel{\eqref{s6-LGS-4a}}\Rightarrow\qquad\vartheta=0\,.
\end{equation*}
That is, there does not exist a vector $0\neq\vartheta\in\dom(s)$, which is orthogonal to
$\dom(\hat{s})$ with respect to $\skal.._t$.
This is a contradiction to our above assumption that the closure $\overline{\hat{s}}$ of the form
$\hat{s}$ of \textsc{part\,(b)} does not agree with the closed, positive form $s$ from \textsc{part\,(a)}.

Our summary in other words: It holds $\overline{\hat{s}}=s$, with $\dom(\overline{\hat{s}})=\dom(s)$,
for the closure of the form $\hat{s}$, or equivalently, $s$ is the smallest closed form extension of the form $\hat{s}$.

All other support cases are proven analogously.

However, for the supports 
(a)--(b), (b)--(b), and (c)--(c) one may arrive faster at the aim $\vartheta=0$ with the following
argumentation:
Remark first that for (a)--(b) and (b)--(b) it is $\skal{1}{\vartheta'}=\int_0^1\vartheta'(x)\,dx=\vartheta(1)-\vartheta(0)=0$
because of the BV $\vartheta(0)=0=\vartheta(1)$ for the orthogonal $\vartheta\in\dom(s)$.
Then search for all polynomials $p(x)$ up to degree $4$, which fulfill the associated four BV.
Of course $p\in\dom(\hat{s})$.
Inserting  $p$ into the orthogonality relation \eqref{s6-orthogonal-rel} in the form
\begin{equation*}
0=\skal{p''''}{\vartheta}+\skal{p'}{1}\skal{1}{\vartheta'} + \skal{p}{1}\skal{1}{\vartheta}\,,
\end{equation*}
and noting that $\skal{p''''}{\vartheta}=c\skal{1}{\vartheta}$ with some constant $c\in\dC$, then one arrives at
$\skal{1}{\vartheta'}=0=\skal{1}{\vartheta}$, simplifying
\eqref{s6-eq-eta-sd=} to $\vartheta''(x)=a+bx$, and thus we end up with a simpler system of linear equations.

\end{small}
%
%

%

\end{sloppypar}
\end{document}